\documentclass[%
 aip,
 amsmath,amssymb,
 reprint,%
]{revtex4-1}

\usepackage{graphicx}%
\usepackage{dcolumn}%
\usepackage{bm}%

\usepackage[utf8]{inputenc}
\usepackage[T1]{fontenc}
\usepackage{mathptmx}

\usepackage{xcolor}

\usepackage{array}
\newcolumntype{P}[1]{>{\centering\arraybackslash}p{#1}}

\begin{document}

\preprint{AIP/123-QED}

\title{Towards empirical force fields that match experimental observables}

\author{Thorben Fr\"ohlking}
\affiliation{Scuola Internazionale Superiore di Studi Avanzati, via Bonomea 265, 34136 Italy}%
\author{Mattia Bernetti}
\affiliation{Scuola Internazionale Superiore di Studi Avanzati, via Bonomea 265, 34136 Italy}%
\author{Nicola Calonaci}
\affiliation{Scuola Internazionale Superiore di Studi Avanzati, via Bonomea 265, 34136 Italy}%
\author{Giovanni Bussi}
 \email{bussi@sissa.it}
\affiliation{Scuola Internazionale Superiore di Studi Avanzati, via Bonomea 265, 34136 Italy}%

\date{\today}%

\begin{abstract}
Biomolecular force fields have been traditionally derived based on a mixture of
reference quantum chemistry data and experimental information obtained on
	small fragments. However, the possibility to run extensive molecular dynamics simulations on larger systems
achieving ergodic sampling is paving the way to directly using such simulations along with solution
experiments obtained on macromolecular systems. Recently, a number of methods have been introduced
to automatize this approach. Here we review these methods,
highlight their relationship with machine learning methods,
and discuss the open challenges in the field.

\end{abstract}

\maketitle

\section{Introduction}

Classical molecular dynamics (MD) simulations at the atomistic scale offer a unique opportunity to
model the conformational dynamics of biomolecular systems. Being able to reveal mechanisms at spatial and temporal scales 
that are difficult to observe experimentally, MD simulations are often seen as a computational microscope.\cite{dror2012biomolecular}
In the past years, they have been applied to study problems ranging from protein folding \cite{lindorff2011fast} and aggregation \cite{baftizadeh2012multidimensional}
to RNA-protein interactions,\cite{perez2015atp,krepl2015can}
transmembrane proteins dynamics,\cite{arkhipov2013architecture} and full
viruses,\cite{freddolino2006molecular}
bacteria,\cite{yu2016biomolecular} or organelles.\cite{singharoy2019atoms}
The capability of MD simulations to reproduce and predict experimental results is
limited by the statistical errors arising from the finite length of simulations and
by the systematic errors resulting from the inaccuracies of the underlying models.
Interactions are often modeled using empirically parametrized force fields that allow timescales
of the order of the microsecond to be routinely simulated.
Importantly, the two sources of error mentioned above are deeply intertwined,
because only systematic errors that are larger than statistical errors
can be detected by comparison with reference experimental results.
Indeed, in the past 20 years, the use of special purpose hardware,\cite{shaw2014anton}
optimized software,\cite{case2005amber,abraham2015gromacs} and enhanced sampling methods,\cite{valsson2016enhancing,camilloni2018advanced} 
has significantly reduced the statistical errors, thereby allowing
force fields inaccuracies to be detected and largely alleviated.
In spite of this, empirical force fields are still far from perfect and
in some cases are poorly predictive. For instance, it is not trivial
to have force fields capable of simultaneously describing correctly folded,
disordered, single-chain proteins or protein complexes,\cite{robustelli2018developing,piana2020development}
to correctly predict RNA structure from sequence-only information
across a wide range of structural motifs,\cite{sponer2018rna}
or to reproduce experimental kinetics in ligand-receptor systems.\cite{capelli2020accuracy}

Solution experiments are optimally suited for validation of force fields, since they provide information
about transiently populated structures as well, and they have traditionally been used in this sense.
Nevertheless, several approaches have enabled solution experiments to be used directly during force-field
fitting, together with available quantum chemistry data.
The aim of this perspective is to review these approaches,
highlight their relationship with machine learning methods,
and discuss the open challenges in the field.

\section{Empirical force fields: bottom up or top down?}

We will use here as paradigmatic examples some of the  force fields that are most used for simulating biomolecular
systems, namely AMBER,\cite{cornell1995amber} CHARMM,\cite{mackerell1995charmm} OPLS,\cite{jorgensen1988opls} and GROMOS.\cite{oostenbrink2004gromos}
All the mentioned force fields share a common functional form, including bond stretching,
angle potentials, torsional potentials, Lennard-Jones, and electrostatic interactions:
\begin{multline}
E= \sum_{bonds} \frac{1}{2}k_b(r-r_0)^2 + \sum_{angles} \frac{1}{2}k_a(a-a_0)^2 + \\ \sum_{torsions} \sum_n\frac{V_n}{2}(1+\cos(n\phi-\delta)) +
\\ \sum_{LJ}4 \epsilon_{ij}\left(\left(\frac{\sigma_{ij}}{r_{ij}}\right)^{12} - \left(\frac{\sigma_{ij}}{r_{ij}}\right)^6\right) + \sum_{electrostatics} \frac{q_iq_j}{r_{ij}} 
\label{eq:ff}
\end{multline}
The parameters ($k_b; r_0; k_a;a_0; V_n; \delta; \sigma; \epsilon; q $) are derived from small fragments in advance and depend on the atom type and its chemical environment.
Polarizable force fields (such as AMOEBA \cite{shi2013polarizable} and a variant of CHARMM \cite{patel2004charmm}),
reactive force fields (such as ReaxFF \cite{senftle2016reaxff}), and
semi-empirical methods (such as DFTB \cite{seifert2012density}) have different functional forms
but similar considerations can be applied.
The parameters in Eq.~\ref{eq:ff}  are derived with a variety of different procedures that depend on the
specific force field and are summarized in Table \ref{table-ff}.
In particular, some of the parameters are typically derived
from quantum chemistry calculations performed at a varying level of accuracy, in a bottom-up spirit.
Other parameters are instead derived from experimental data, either using spectroscopy experiments,
databases of crystallographic structures, or other gas-phase or solution-phase experiments, in a top-down
spirit.

\begin{table*}
\centering
	\caption{Collection of commonly used force fields and
method used in their original version for obtaining the respective parameter sets
(reference to the original paper is reported for each force field family).
A more detailed table is reported in Supplementary Material.
\label{table-ff}
}
\begin{tabular}{ P{2cm}|P{4cm}| P{4cm} |P{4cm}| P{2cm}  }
& \textbf{AMBER} \cite{cornell1995amber} & \textbf{CHARMM} \cite{mackerell1995charmm} & \textbf{OPLS} \cite{jorgensen1988opls} & \textbf{GROMOS} \cite{oostenbrink2004gromos}\\
\hline
Bond &  Experiments & Experiments + Ab initio &  AMBER parameters &    Experiments \\
\hline
Bend & Experiments &  Experiments + Ab initio  &  AMBER parameters &  Experiments \\
\hline
Torsion & Experiments + Ab initio & Experiments + Ab initio  &  AMBER parameters & Ab initio \\
\hline
LJ  & Monte Carlo liquid simulations + OPLS parameters  & Experiments + Ab initio  &  Experiments + Ab initio + Monte Carlo liquid simulations   &    Experiments  \\
\hline
Charges & Ab initio & Experiments + Ab initio &  Experiments + Ab initio + Monte Carlo liquid simulations &  Experiments \\
\end{tabular}
\end{table*}

\begin{figure}
\includegraphics[width=0.85\columnwidth]{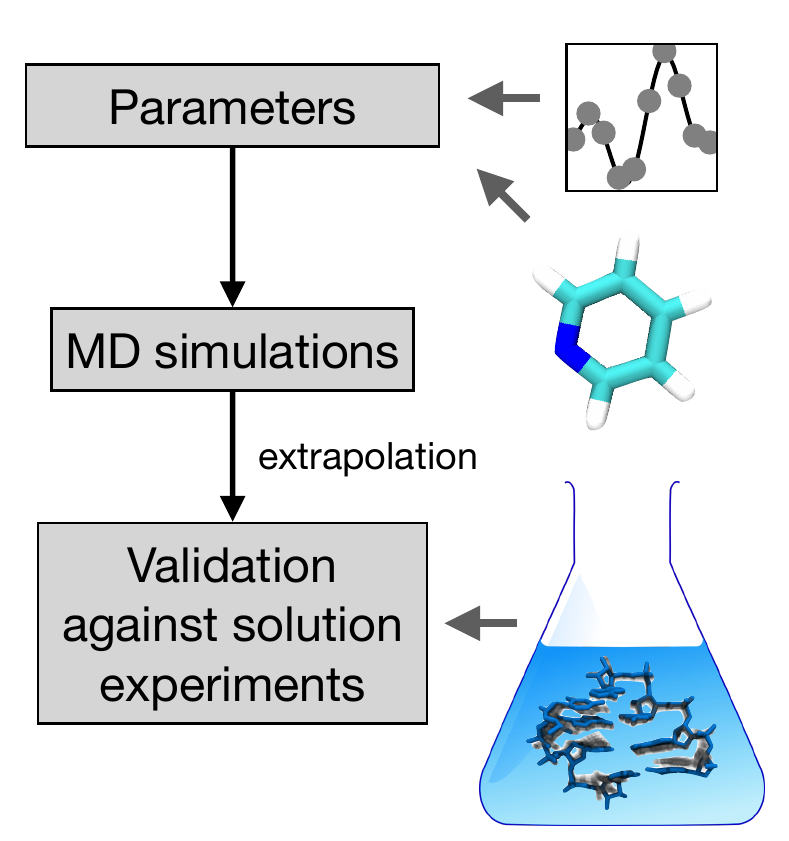}
\caption{
Traditional procedure used for force-field parametrization.
Parameters are obtained from calculations or experiments on small molecules or fragments.
Simulations are then validated for their capability to maintain the native structure
of a macromolecule or against solution experiments. Since fitting and validation
are done on different types of systems, there is a large risk associated to extrapolation.
\label{fig_fit1}
}
\end{figure}

\begin{figure}
\includegraphics[width=0.9\columnwidth]{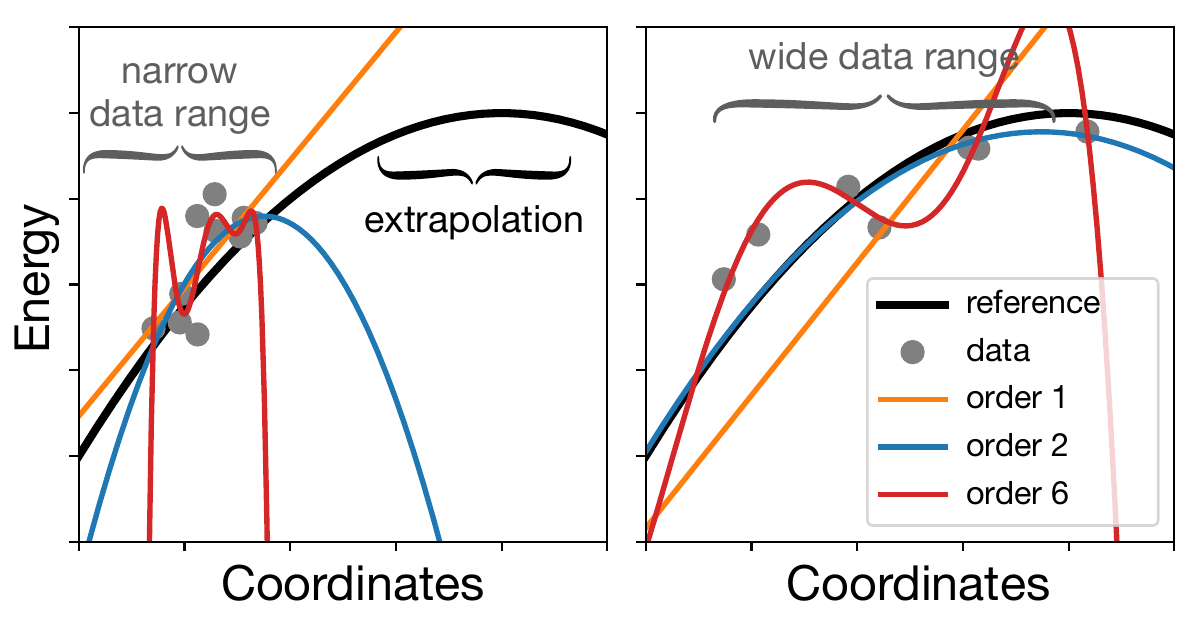}
\caption{
Typical errors observed when fitting a function and extrapolating.
The horizontal axis represents a configurational coordinate (\textit{e.g.}, a dihedral angle) and the vertical axis an
observable that is used for fitting (\textit{e.g.}, the energy of the system).
The true function is shown as a solid line, and the available reference data are shown as grey points.
Lines fitted on the reference data using polynomials of increasing order (order 1, 2, and 6)
are shown in colors.
Data are collected on a narrow (left panel) or wide (right panel) range of configurations.
The simple model (order 1) represents a force field with too few parameters or with an incorrect functional form.
When fitted on a narrow range of configurations (left panel) it reproduces well the true function.
However, it fails the extrapolation to the right part of the graph.
When fitted on a wide range of configurations (right panel) 
the intrinsic limited transferability of the model emerges from the error observed on the fitted points.
The complex model (order 2) represents a force field with more parameters and a more physical functional form,
since also the reference curve is an order 2 polynomial.
When fitted on a narrow range of configurations (left panel) it can lead to significant overfitting.
Conversely, when fitted on a wide range of configurations (right panel), it reproduces well the true function
on the entire range of configurations.
The highest order polynomial (order 6) overfits the reference data in both cases.
\label{fig_fit2}
}
\end{figure}
One of the factors impacting the reliability of a force field is the 
accuracy of the employed reference data.
For instance, a force field fitted purely on quantum chemistry data cannot provide results that
are more accurate than the reference method. However, this limit can be surpassed if multiple sources of data
are combined.
As an additional and perhaps even more important source of error,
one should take into account that
reference data used in force-field fitting, either computational or experimental ones, are obtained studying systems
that are necessarily not identical to those that one wants to simulate later (see Fig.~\ref{fig_fit1}).
For instance, torsional parameters and partial charges in the AMBER force field are traditionally obtained
using quantum chemistry calculations in small fragments of up to a few dozen atoms,
typically including a couple of aminoacids, but are later used to simulate oligopeptides or full protein domains.
Similarly, Lennard-Jones parameters in the OPLS force field are obtained from vaporization calorimetry of pure organic liquids such as tetrahydrofuran, pyridine or benzene, but then applied to cases where the analyzed compounds are only portions of
a sugar, nucleobase, or aminoacid respectively. 
These parameters have not been changed in more recent OPLS versions.
The reliability of a force field when used in a context different from the one in which it was parametrized
depends on the transferability of the functional form in Eq.~\ref{eq:ff}
(see Fig.~\ref{fig_fit2}).
Given the very large gap between the size and complexity of the systems used for parameter fitting
and the systems to which force fields are applied,
it appears almost a miracle that current force fields
are, for instance, capable of correctly identifying the folded state of a protein.\cite{lindorff2011fast}

It is interesting to look at a few anecdotal examples to better understand how this is possible.
The traditional AMBER force field for nucleic acids has been used for several years before it was realized
that sufficiently long simulations could lead to a transition to experimentally unobserved
rotamers in the $\alpha$ and $\gamma$ torsions of DNA backbone.\cite{varnai2004dna,perez2007refinement}
Following this empirical observation, a joint effort of several groups lead to the
parmbsc0 reparameterization of DNA backbone,\cite{perez2007refinement} where the parameters corresponding to these two
torsional angles were fitted against quantum chemistry calculations.
A similar episode occurred later with the $\chi_{OL3}$ corrections, derived to counteract the
occurrence of ladder-like structures in RNA.\cite{zgarbova2011refinement}
In the CHARMM force field for proteins, one of the most important additions after its initial development
has been the introduction of empirical corrections maps (CMAP) 
\cite{mackerell2004extending}, that deviate from the functional form of Eq.~\ref{eq:ff}
by the presence of coupling terms between consecutive torsional angles.
These corrections were fitted on quantum chemistry data,
but required also a heuristic adjustment to fix the typical values of torsional angles
in $\alpha$-helical and $\beta$-sheet regions.
As a further example, empirical adjustments of the AMBER and CHARMM force fields were
performed respectively in Ref.~\onlinecite{best2009optimized} and in Refs.~\onlinecite{piana2011robust,best2012optimization},
where solution data on short oligopeptides were used to optimize backbone dihedrals
so as to reproduce helix-coil transitions.

A general trend that can be seen is that
experimental data on macromolecular systems (\textit{e.g.}, nucleic acids
duplexes or protein domains) are typically used for validation, whereas the parameters
are fitted on either theoretical or experimental information available for much smaller systems.
Nonetheless, the observation of failures in macromolecular systems is the only way
to detect which precise parameters should be corrected.
The last three mentioned works, \cite{best2009optimized,piana2011robust,best2012optimization} instead, report direct
fitting of parameters on simulations of short oligomers.

\section{Recent approaches for fitting force field parameters on experimental data}

\begin{figure}
\includegraphics[width=\columnwidth]{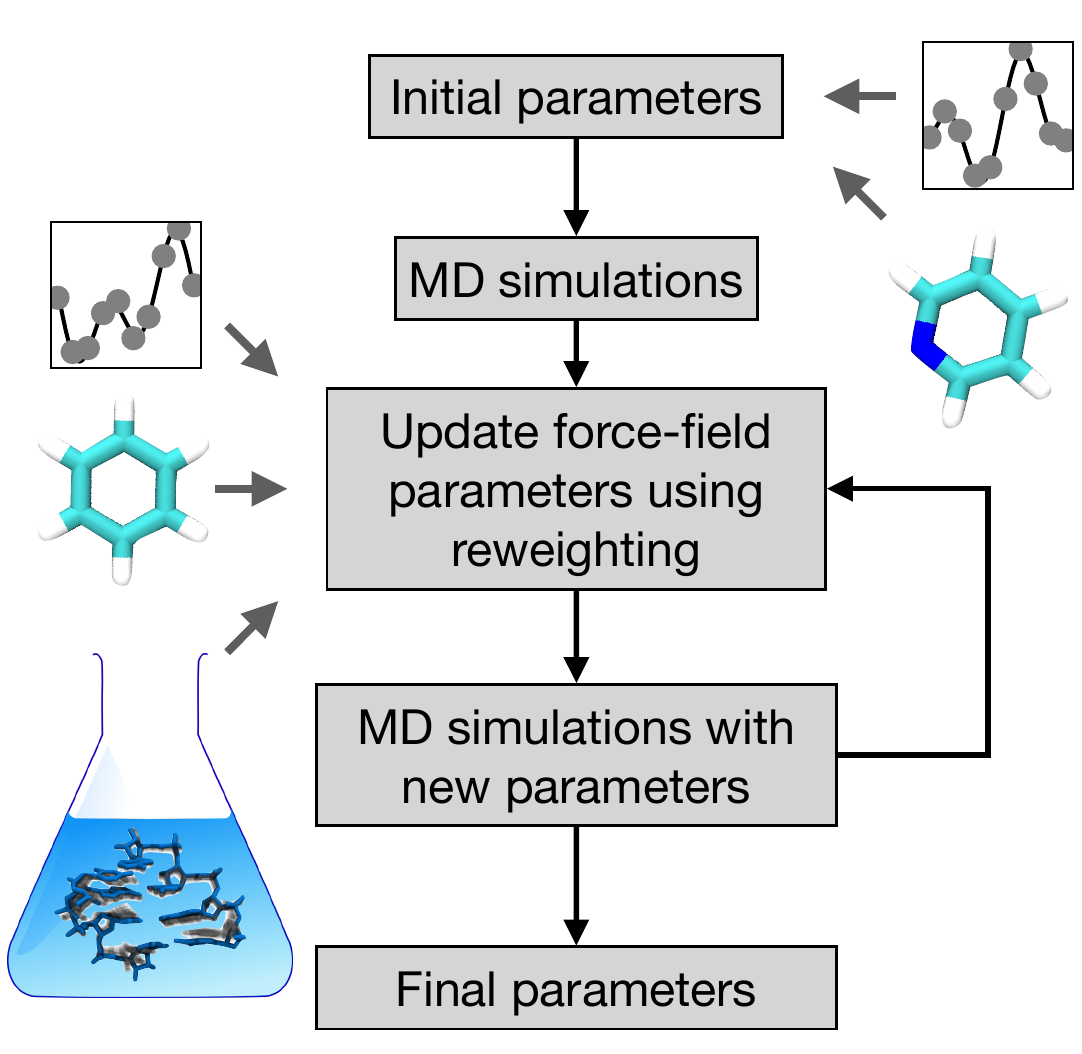}
\caption{
Schematic representation of a force-field fitting procedure using experimental data on macromolecular systems.
Initial parameters are tuned based on both quantum chemistry data and experimental
data on small systems (\textit{e.g.}, individual residues). Molecular dynamics simulations are then performed on macromolecular systems.
Reweighting is used to optimize force field parameters in order
to maximize the agreement with a set of available data including experiments on macromolecular systems.
In principle, this second stage might include also quantum chemistry data and experimental data on small systems.
Even when not explicitly used at this stage, the initial set of quantum chemistry and experimental data
is still playing a role for all the parameters that are not further adjusted. Even on the adjusted
parameters, the information about the initial force field remains present if regularization terms are included.
\label{fig_scheme}
}
\end{figure}

A number of approaches have been introduced to allow fitting force fields directly on experimental data
taken on macromolecular systems rather than on small fragments, all of them following a flowchart similar to the one
illustrated in Fig.~\ref{fig_scheme}.
Since solution experiments often report results that are averaged over an ensemble of copies of the same
molecule, these methods are typically designed to enforce ensemble averages rather than instantaneous values.
Norgaard \textit{et al.}~\cite{norgaard2008experimental} introduced an approach where a force field
is iteratively refined until agreement with experiment is obtained. At each iteration,
a simulation is
performed and the force field parameters are optimized
by assigning new weights to the visited conformations. Thus, 
through such reweighting procedure, one can predict what result
would be obtained using these slightly modified parameters.
At some point,
when the refined force field and the initial one become too different, it is necessary to iterate the procedure
performing a new simulation.
The method was applied to the refinement of a coarse-grained model of a protein
and fitted against paramagnetic relaxation enhancement experiments.
The same method was later used to choose the parameters of an implicit-solvent model
against reference all-atom simulations.\cite{bottaro2013variational}
Li \textit{et al.}~\cite{li2011iterative} showed how to refine an all-atom protein force field using chemical shifts and full-length protein simulations.
A common trait of all these methods is that even small changes in force field parameters can make the resulting ensemble
very different from the original one making the reweighting procedure less accurate. In Ref.~\onlinecite{li2011iterative},
a local reweighting procedure was introduced to alleviate this issue. This procedure is based on the heuristic observation that
the ensemble of conformations accessible to a residue is maximally affected by the parameters used for that residue and, to a lesser extent,
by the parameters used for the other (possibly identical) residues. Since this is an approximation, a
subsequent simulation performed with the corrected force field was necessary to validate the modification.
Refs.~\onlinecite{wang2012systematic,wang2014building} used a similar automatic procedure to optimize water models.
Interestingly, they realized that a straightforward fitting procedure might lead to overfitting and showed
how a regularization term can be included in order to alleviate this issue.
Finally, Cesari \textit{et al.}~\cite{cesari2019fitting} introduced a procedure to refine atomistic force fields where heterogenous systems
and types of experimental data are used to refine the AMBER RNA force field.
Enhanced sampling techniques are employed by the authors to ergodically sample the conformational space for a number of RNA tetramers and hairpin loops,
and a regularization term is used in the fitting scheme to maintain the refined force field close to the initial one.
The weight of the regularization term is chosen with a cross-validation procedure
aimed at maximizing the transferability of the parameters.

\begin{figure}
\includegraphics[width=\columnwidth]{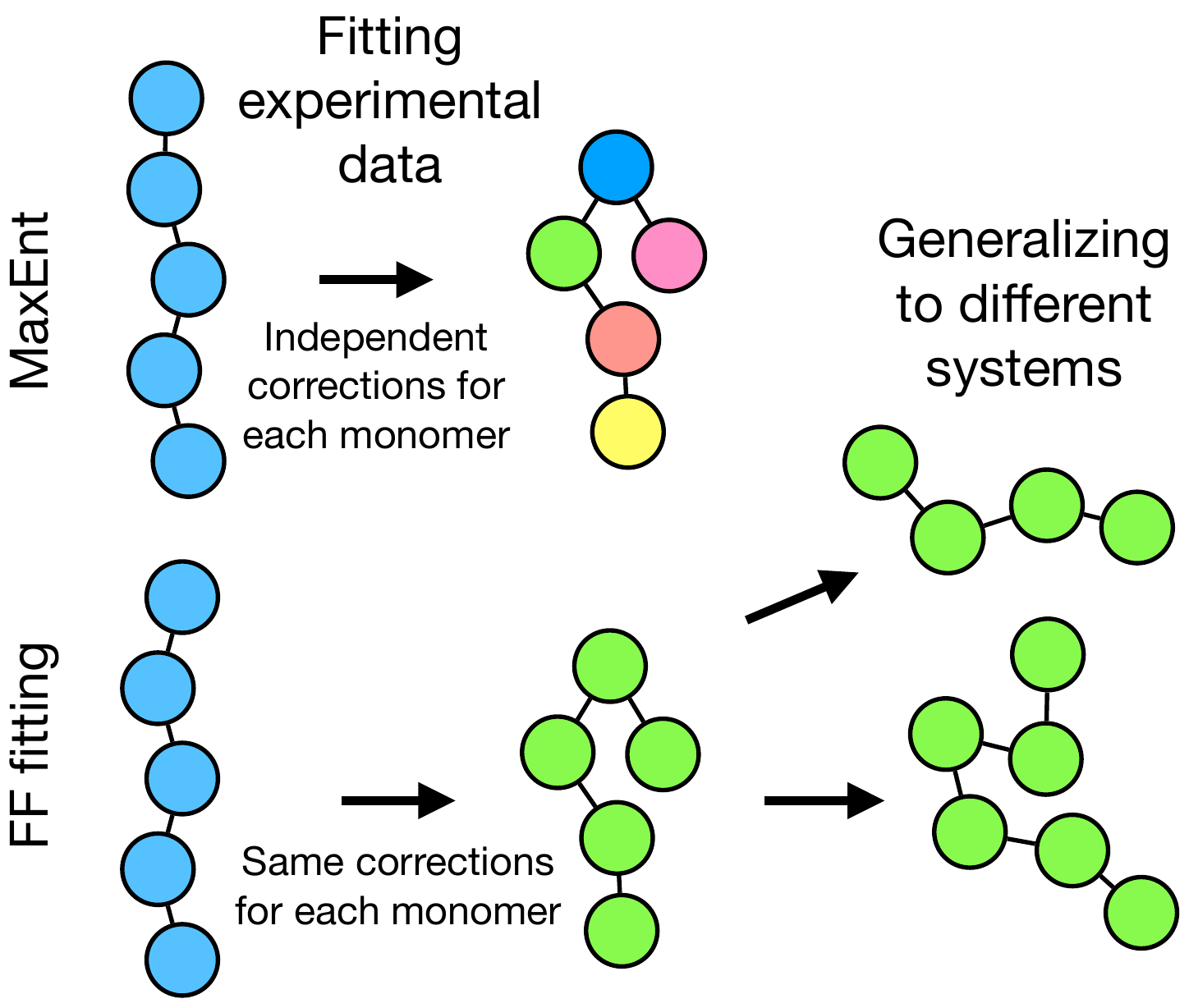}
\caption{
Difference between maximum entropy and force-field fitting procedures.
When using the maximum entropy principle to enforce agreement between simulation and experiment,
one free parameter is used for each data point. As a consequence, different chemically equivalent units
might be treated differently.
This does not allow the corrections to be transferred to other molecules,
for which new experimental data would be required.
When using force-field fitting procedures, instead, all chemically equivalent
units are treated in the same manner. This allows the derived parameters to be generalized to other molecules
where the same units are used as building blocks.
}
\label{fig-maxent}
\end{figure}
It is important to recognize the difference between the mentioned approaches, which are meant to generate transferable force-field parameters,
and methods meant to 
improve the agreement with experiment for a specific system for which data are available.\cite{bonomi2017principles}
This second class includes a variety of approaches such as Bayesian schemes\cite{hummer2015bayesian,bonomi2016metainference} and methods based
on the maximum entropy principle.\cite{pitera2012use,cesari2018using}
In the maximum entropy formalism the number of free parameters is equal to the number of experimental datapoints.
For instance, in a homogenous polymer, each of the monomers will feel a different correction that makes its structure as compatible as
possible with experiments.
Since the number of parameters is very high, regularization methods can be used and tuned
with a cross-validation procedure (see, \textit{e.g.}, Ref.~\onlinecite{bottaro2018conformational}).
In addition, if a polymer of a different length needs to be simulated, new experimental data should be obtained.
In force-field fitting procedures, the chemical structure of the investigated molecule is \textit{a priori} used to
reduce the number of parameters. For instance, in a homogenous polymer, each of the monomers will feel the same correction
(although perhaps terminal monomers might be treated differently\cite{mlynsky2020fine}).
On the one hand, this allows to encode a large amount of information in the specific choice of the functional form employed.
This type of information is similar to the one that is included when atoms are classified in types in order to obtain
their parameters.\cite{mobley2018escaping}
On the other hand, it significantly reduces the number of parameters potentially making the resulting force field transferable.
Ref.~\onlinecite{cesari2016combining} used a hybrid approach were maximum entropy restraints were used but
kept by construction constant across chemically equivalent parts of the system.
For a recent comparison of approaches taken from both classes, see Ref.~\onlinecite{orioli2019learn}.

Besides the discussed systematic approaches, that report methodological improvements
aimed at optimizing parameters based on experimental data,
a number of recently developed force fields include terms that were manually adjusted based on
the result of MD simulations on systems of different complexity and their capability to reproduce
experimental data.
For instance, Refs.~\onlinecite{robustelli2018developing,best2009optimized,piana2011robust,best2012optimization}
reported optimizations
of parameters based on the solution properties of oligopeptides.
The atomic radii of the AMBER ff15ipq force field were chosen so as to provide correct salt-brigde interactions.\cite{debiec2016further}
Finally, two recent variants of the AMBER RNA force field contain corrections on hydrogen bonds obtained scanning a series
of parameters and minimizing the discrepancy with solution experiment for RNA oligomers.\cite{kuhrova2019improving,mlynsky2020fine}

\section{The machine learning lesson: how to avoid overfitting}

\begin{figure}
\includegraphics[width=\columnwidth]{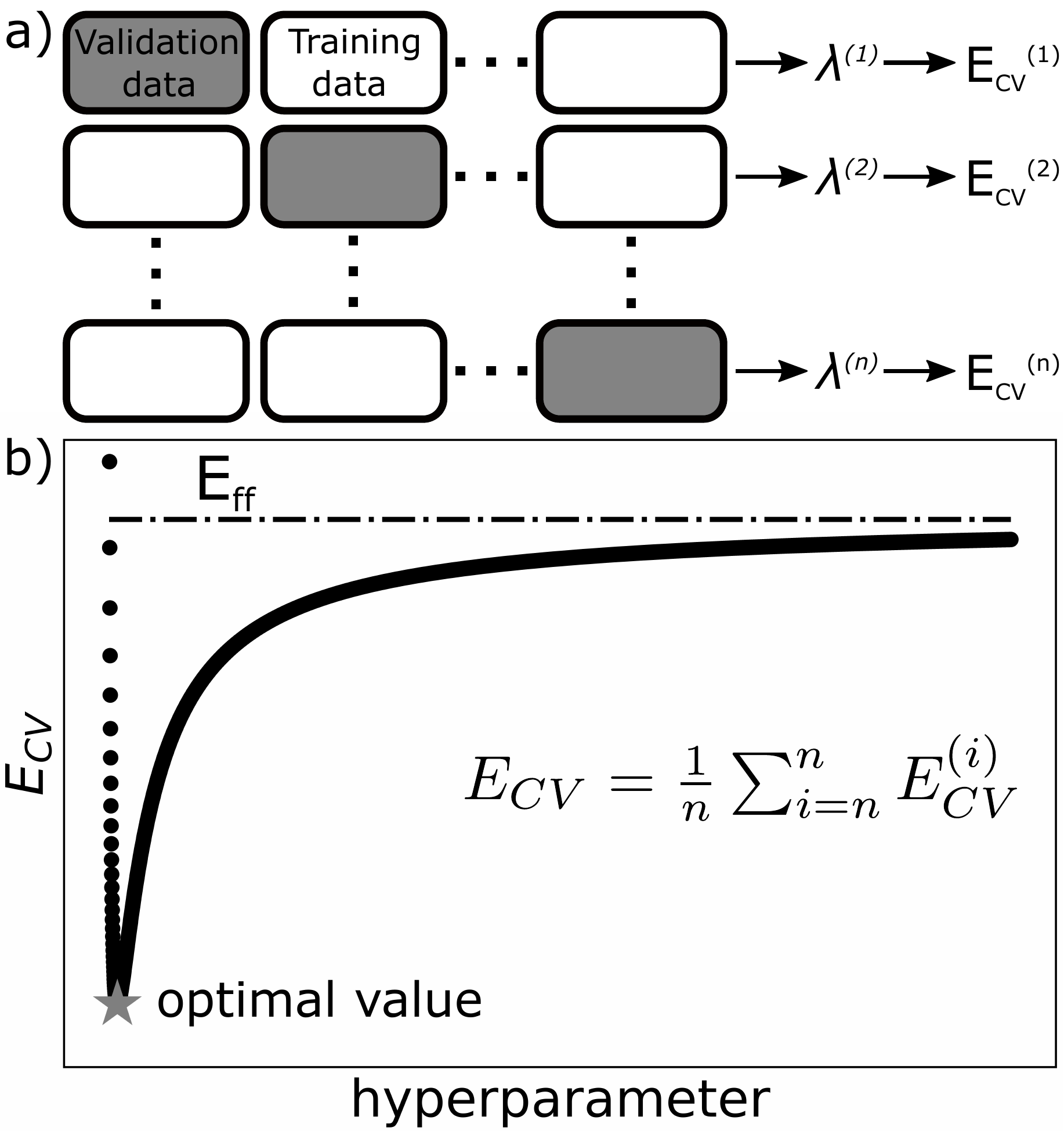}
\caption{Cross-validation can be used to decrease overfitting and allow more generalizable force-field improvements.
a) In $n$-fold cross validation, the data set is randomly split into $n$ blocks of equal size. A single block of data
is left out and the parameters are trained on the remaining $n-1$ blocks.
This is repeated once for each block, yielding multiple sets of trained parameters $\lambda^{(1)},\dots,\lambda^{(n)}$.  
The error $E_{CV}^{(k)}$ obtained with parameters $\lambda^{(k)}$ in reproducing the $k$-th block is then computed,
and the average over the $n$ results is the cross-validation error ($E_{CV}$).
Leave-one-out cross validation is a special case where each block contains a single data point,
whereas in leave-$p$-out cross validation all possible subsets of $p$ data are left out.
b) Hyperparameters controlling model complexity (such as regularization coefficients) are then chosen
so as to minimize the cross-validation error.
\label{fig-leave-one-out}
}
\end{figure}

Overfitting is a ubiquitous problem when fitting procedures are done in a blind manner.
The prototypical cases are machine learning and related algorithms where functions of arbitrary complexity,
supported by no or little physical understanding, are used to fit empirical data.
The machine learning community has thus developed a number of tools that can be used to avoid
or at least alleviate this issue.

Many different machine learning techniques exist and are typically based on a common framework.\cite{mehta2019ml} The basic ingredient is a dataset made up of a
set of independent variables ($\mathbf{X}$, samples) and a set of dependent variables ($\mathbf{Y}$, labels).
Next, a set of models is proposed to map $\mathbf{X}$ into $\mathbf{Y}$ with best accuracy. A model is defined by a set of parameters plus a set of hyperparameters. 
This splitting is guided by computational convenience such that inference can be approached in a multi-level fashion: typically, model parameters 
are found by solving an optimization problem at fixed hyperparameters, that on the other hand are preferably scanned over a discrete scale. 
This double approach is more easily understood when another basic ingredient of machine learning is introduced, that is the cost function. 
The cost function is used to estimate the performance of a model, and while it is usually a continuous function of the model parameters, 
it can have a non-trivial dependence on the hyperparameters. For example, the set of hyperparameters can include the architecture of the model, 
the optimization algorithm used to find the optimal model parameters, the functional form of the cost function itself, etc. 
Similar choices also need to be taken in force field fitting, as discussed in detail in the next Section.

Since the sets of parameters and hyperparameters defining models are fitted against a finite set of examples $\{\mathbf{X},\mathbf{Y}\}$, overfitting can easily occur.
In the limit of fitting on an infinite amount of data, the only limitation of a model would be determined by its complexity.
In this limit, a too simple model would \textit{underfit} the data, leading to a \textit{bias} in the result. This bias can be decreased by increasing the model complexity.
But since in general we deal with datasets of finite size, increasing the complexity of the model would result in a large contribution to the error
(\textit{variance}) due to the sampling.
A too complex model would \textit{overfit} the data, thus having a seriously low performance on new independent data.    

The search for the model with the optimal tradeoff between bias and variance (i.e. between under- and over-fitting) follows two directions. 
One is to split the dataset into a training and a cross-validation set, prior to analysis. Model parameters are fitted against 
data in the training set, and afterwards the optimized model is validated against the validation set data not included in the training procedure. 
This procedure is usually referred to as cross-validation (Fig.~\ref{fig-leave-one-out}),
and depending on how the cross-validation set is built can be referred to as leave-one-out, leave-$p$-out,
or $n$-fold cross validation.
The other is to reduce the risk of overfitting by means of regularization techniques, 
the most common consisting in adding terms to the cost function that prevent the model parameters from reaching values extremely adapted to the 
dataset. 
This comes at the cost of increasing the number of hyperparameters (\textit{e.g.}, the relative size of training and cross-validation sets, their composition, 
coefficients of regularization terms, etc.) that continue to be affected by risk of overfitting. Even if a close solution to this problem is not 
established yet, overfitting should be taken into account for each level of inference (for both parameters and hyperparameters). 
The most straightforward way to deal with this multi-level risk of overfitting is to \textit{a priori} split the dataset into 
three subsets: in addition to the standard training and cross-validation subsets, an independent test set is introduced. The training set
is used to fit the optimal values of parameters at fixed hyperparameters; optimal hyperparameters are then fitted against
the cross-validation set. Eventually, the performance of the model defined by the optimal parameters and hyperparameters is
evaluated on the test set.  
A more robust approach consists in nested cross-validation \cite{cawley2010over}, in which parameters and hyperparameters are optimized
on a single dataset, but the criterion used to optimize model parameters (training) is different from the optimization criterion used 
for hyperparameters (model selection). Validation of the selected optimized model against new data that, 
importantly, has not been used to adjust neither parameters nor hyperparameters, is best practice in this case as well.

\section{Overfitting in force field development}

\begin{figure}
\includegraphics[width=\columnwidth]{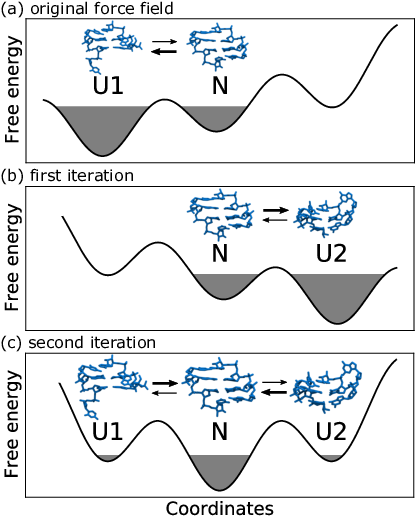}
\caption{
\label{fig-reweighting}
Free-energy landscapes when using reweighting and iterative simulations.
The native state, compatible with experimental information, is in the middle (N).
Two metastable non-native states that, based on experimental information, are supposed to have a low population,
are also shown (U1 and U2).
In the original force field (panel a), the N state is sampled, but the most stable state
is U1. During the reweighting procedure, the force field learns how to improve
the agreement with experiments by disfavoring U1. However, since U2 was never observed in this simulation, there is nothing that prevents
it to be stable when using the refined force field.
Once a simulation is performed with the refined force field (panel b), the state U2
appears with large population leading to disagreement with experiment.
In principle, if a reweighting is performed using only the second simulation, state U1 might appear again with an incorrect population.
Only a reweighting where both simulations are combined, and thus all the possible states can be observed,
is capable of generating a force field that correctly sets N as the global free-energy minimum and U1 and U2
as metastable states with low population (panel c).
}
\end{figure}
Force-field fitting procedures can be interpreted as machine learning methods where the parameters are the optimized coefficients 
and data and labels are a mixture of information obtained from both quantum chemistry calculations and various experimental techniques.
One should thus pay attention to overfitting.
Whenever overfitting occurs, transferability of the force field to a different case might be compromised.
As already discussed, if parameters are only fitted on small systems, their transferability to larger systems might be limited.
The other phenomenon that can be observed in reweighting methods is the subtle overfitting on the analyzed trajectory.
In particular, if parameters are derived to match experimental data by reweighting a trajectory that
is not sufficiently long,
they might not work correctly on another trajectory obtained using the same force field but with different initial conditions.
In addition,
since reweighting schemes can only modulate the weight of states that have been explored but cannot predict the
population of states that have not been observed (see, \textit{e.g.}, Ref.~\onlinecite{rangan2018determination} for a comparison of restraining
and reweighting when used to implement the maximum entropy principle), the only way to detect these problems is to keep the target force field
as close as possible to the original one with some form of regularization and to then perform a new simulation once parameters have 
been optimized (see Fig.~\ref{fig-reweighting}).

\begin{figure}
\includegraphics[width=\columnwidth]{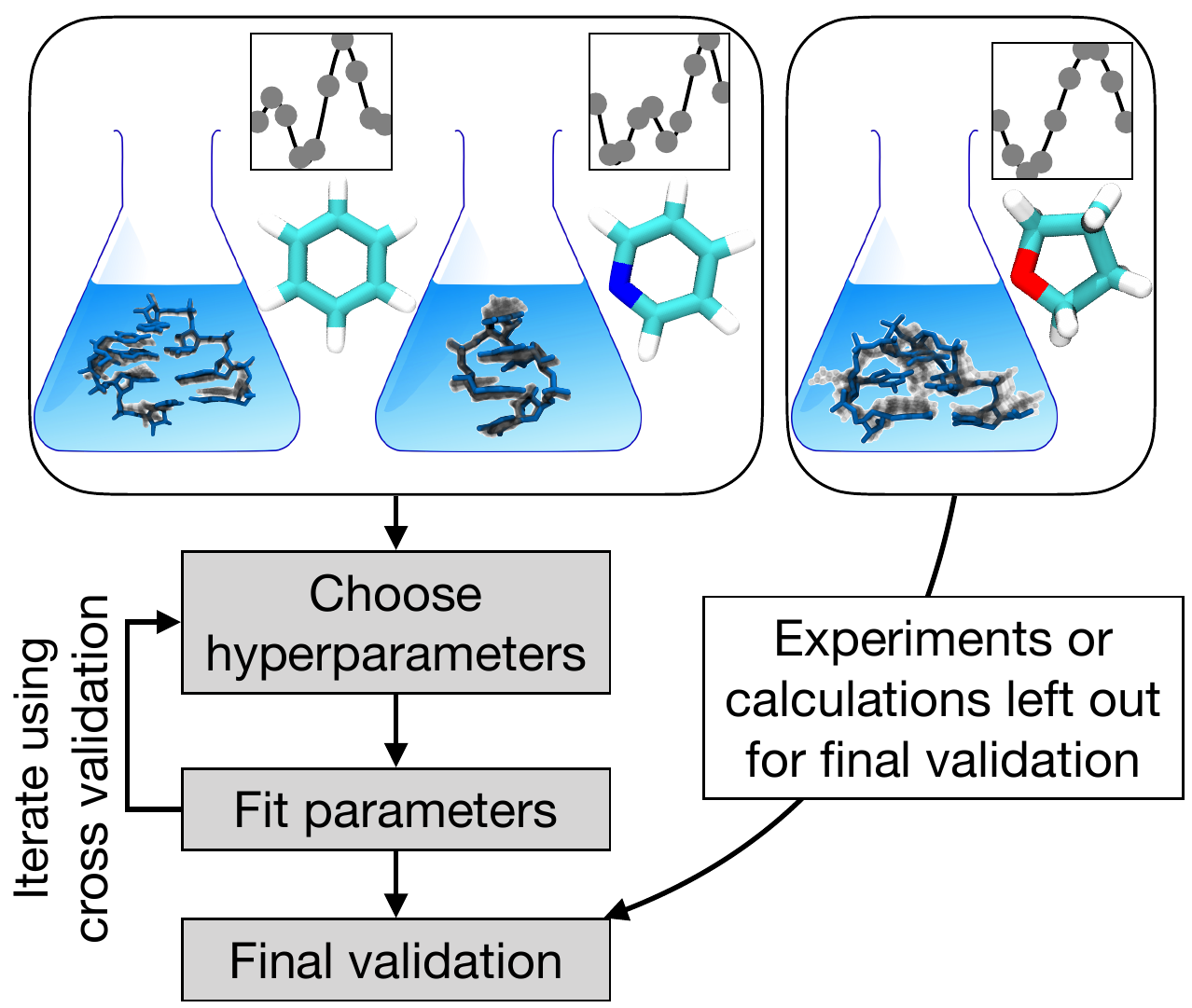}
\caption{
Schematic representation of a training/final-validation procedure.
Force-field fitting can be based on a combination of quantum chemistry data,
experimental data on small systems and experimental data on macromolecular systems.
All data can be used in parameter fitting, and cross-validation (see Fig. \ref{fig-leave-one-out})
can be used to help the choice of hyperparameters. To this end, it is necessary that a separate
set of data, either theoretical or experimental, is left out until the very end of the procedure
to validate the transferability of the model. This separate data should not be used to take
any decision during the fitting procedure, or the information leak might 
make the final validation not truly independent.
\label{fig-final-validation}
}
\end{figure}
Every other decision taken in the path should be included in the list of hyperparameters.
Coefficients controlling regularization terms used in the optimization, that control the relative weight of initial force field and of experimental data,
are naturally considered hyperparameters.
The functional form of the force field itself is a hyperparameter.
The analogous of these hyperparameters in the training of neural network are regularization terms or early stopping criteria
and network architecture, respectively.\cite{goodfellow2016deep}
In force-field fitting, a number of additional hyperparameters might be used whose control might be more or less explicit.
For instance, the so-called forward models used to calculate experimental observables from MD trajectories contain a number of
parameters.
If the training is done to reproduce the energetics of quantum chemistry calculations, the set of structures used for fitting and their
relative weights are to be considered as hyperparameters. Even the precise quantum methods used to compute the total energy might contain a number
of hidden parameters (\textit{e.g.}, the possibility to use either implicit or explicit solvent or the specific method used to solve
the many-body Schr{\"o}dinger equation).

If hyperparameters are chosen \textit{a priori} based on some independent intuition or information,
for instance the fact that a given quantum chemistry method is more accurate than another one,
then this extra information will be encoded in the final result, improving the quality of the resulting
model. However, if hyperparameters are optimized by monitoring the performance of the force field on a specific system,
then this system will implicitly become part of the training set. Thus, the resulting model should be
validated against a separate system (Fig.~\ref{fig-final-validation}).
A practical example would be if different variants of a force field are derived
using three different quantum chemistry methods, then the best method
is chosen checking the stability of the native structure of a specific system using all the derived variants.
Unless there are other independent evidences that the selected quantum chemistry method is better than the other ones,
this choice should be considered as fitted on the specific system and should be then validated on an independent one.
Therefore, as a final remark, all the decisions taken in the process should be critically evaluated in this respect.

\section{Critical issues and open challenges}

The recent works done in adjusting force field parameters including experimental data
suggests that this is a promising field that will lead to important improvements
in the future. There are however a number of critical issues that one should
carefully consider.

First, we suggest that all input data should be considered at the same time,
irrespectively of being obtained from experiment or from quantum chemistry calculations.
Both experimental and quantum chemistry data can indeed be equivalently used for training or for validation.
Importantly, one should consider that different data have different relative errors
and different information content. Data that provide limited information during fitting
will also provide less stringent validations.
Particularly valuable are data obtained on systems as close as possible to those
that one is interested in simulating. Less weight instead should be given to data
obtained in very different conditions (\textit{e.g.}, without solvent) or on systems that are
too simple to be considered as representative (\textit{e.g.}, individual aminoacids or nucleotides).
As an exception to this general rule, one should consider that different types of data
typically give access to the energetics in different portions of the conformational space.
For instance, solution experiments on macromolecular systems are valuable in providing the relative stability of
structures that can be distinguished using some probe. Quantum chemistry calculations
are instead valuable when states are difficult to be distinguished in the experiment,
or when probing rarely visited states (such as transition states).

Reference data should be obtained in conditions as realistic as possible.
One should carefully
consider the conditions in which experiments are carried out,
and prefer experiments performed in conditions that can be reproduced in MD simulations.
Ideally, specific experiments might be designed and performed in order to facilitate force field development,
such as solution experiments on systems small enough to allow ergodic sampling but large enough to provide
transferable information.
When instead basing the fits on quantum chemistry calculations,
one should consider the importance of the solvent. Additionally, errors in the experimental data
should be taken into account. Very important are also errors in the forward models used to connect
structures obtained in MD simulations with experiments, since these errors are often larger than the those of the raw data.
Errors in the quantum chemistry calculations should be quantified as well.

Taking inspiration from the machine learning community, it is fundamental to understand how to
avoid overfitting. In particular, overfitting on specific systems should be avoided, and this can be achieved by including
as heterogenous as possible systems in the dataset. Similarly, overfitting should be avoided on
specific trajectories. To this end, separate validation simulations can be run or robust estimates of
the statistical errors can be pursued.
Regularization terms can be used to tune model complexity thus reducing the impact of overfitting.
Validation should be made on data that are obtained in an as independent as possible manner.

Finally, the current functional form (Eq.~\ref{eq:ff}) might be too limited to be usable
on a wide range of cases. Increasing the complexity of the model might help in this respect.
Complexity can be introduced by physical insight (\textit{e.g.},
explicitly polarizable force fields \cite{patel2004charmm,shi2013polarizable} or
modified Lennard-Jones potentials \cite{li2014taking})
or by blind learning of non-linear models (\textit{e.g.}, neural network potentials \cite{behler2007generalized,smith2017ani}).
Nonetheless, one should keep in mind that, whenever complexity is increased, overfitting has more chance to appear.
In this respect, for a fixed number of parameters, the more physical the functional form is, the less it will tend to overfit.
Interestingly, neural networks are now routinely used to fit bottom up potentials where the training data
can be generated by computational methods and
can then be easily made very abundant.\cite{behler2007generalized,smith2017ani,noe2020machine,gkeka2020machine}
These approaches are however typically designed to be trained on very small systems or chemical groups, and
their applicability to macromolecular systems has not been shown yet.
It is thus still to be seen if neural network potentials can be used fruitfully when force fields are directly fitted on experimental data.

\section*{Supplementary Material}
A table reporting more details about the methods used to obtain parameters in the most common biomolecular force field.

\begin{acknowledgments}
Sandro Bottaro,
Lillian Chong
and Andreas Larsen
are acknowledged for carefully reading the manuscript and providing useful suggestions.
Max Bonomi is greatly acknowledged for carefully reading the manuscript and reporting
enlightening feedbacks.
\end{acknowledgments}

\section*{Data availability}
Data sharing not applicable --- no new data generated.

\bibliography{main}%

\begin{thebibliography}{60}%
\makeatletter
\providecommand \@ifxundefined [1]{%
 \@ifx{#1\undefined}
}%
\providecommand \@ifnum [1]{%
 \ifnum #1\expandafter \@firstoftwo
 \else \expandafter \@secondoftwo
 \fi
}%
\providecommand \@ifx [1]{%
 \ifx #1\expandafter \@firstoftwo
 \else \expandafter \@secondoftwo
 \fi
}%
\providecommand \natexlab [1]{#1}%
\providecommand \enquote  [1]{``#1''}%
\providecommand \bibnamefont  [1]{#1}%
\providecommand \bibfnamefont [1]{#1}%
\providecommand \citenamefont [1]{#1}%
\providecommand \href@noop [0]{\@secondoftwo}%
\providecommand \href [0]{\begingroup \@sanitize@url \@href}%
\providecommand \@href[1]{\@@startlink{#1}\@@href}%
\providecommand \@@href[1]{\endgroup#1\@@endlink}%
\providecommand \@sanitize@url [0]{\catcode `\\12\catcode `\$12\catcode
  `\&12\catcode `\#12\catcode `\^12\catcode `\_12\catcode `\%12\relax}%
\providecommand \@@startlink[1]{}%
\providecommand \@@endlink[0]{}%
\providecommand \url  [0]{\begingroup\@sanitize@url \@url }%
\providecommand \@url [1]{\endgroup\@href {#1}{\urlprefix }}%
\providecommand \urlprefix  [0]{URL }%
\providecommand \Eprint [0]{\href }%
\providecommand \doibase [0]{http://dx.doi.org/}%
\providecommand \selectlanguage [0]{\@gobble}%
\providecommand \bibinfo  [0]{\@secondoftwo}%
\providecommand \bibfield  [0]{\@secondoftwo}%
\providecommand \translation [1]{[#1]}%
\providecommand \BibitemOpen [0]{}%
\providecommand \bibitemStop [0]{}%
\providecommand \bibitemNoStop [0]{.\EOS\space}%
\providecommand \EOS [0]{\spacefactor3000\relax}%
\providecommand \BibitemShut  [1]{\csname bibitem#1\endcsname}%
\let\auto@bib@innerbib\@empty
\bibitem [{\citenamefont {Dror}\ \emph {et~al.}(2012)\citenamefont {Dror},
  \citenamefont {Dirks}, \citenamefont {Grossman}, \citenamefont {Xu},\ and\
  \citenamefont {Shaw}}]{dror2012biomolecular}%
  \BibitemOpen
  \bibfield  {author} {\bibinfo {author} {\bibfnamefont {R.~O.}\ \bibnamefont
  {Dror}}, \bibinfo {author} {\bibfnamefont {R.~M.}\ \bibnamefont {Dirks}},
  \bibinfo {author} {\bibfnamefont {J.}~\bibnamefont {Grossman}}, \bibinfo
  {author} {\bibfnamefont {H.}~\bibnamefont {Xu}}, \ and\ \bibinfo {author}
  {\bibfnamefont {D.~E.}\ \bibnamefont {Shaw}},\ }\bibfield  {title} {\enquote
  {\bibinfo {title} {Biomolecular simulation: a computational microscope for
  molecular biology},}\ }\href@noop {} {\bibfield  {journal} {\bibinfo
  {journal} {Annu. Rev. Biophys.}\ }\textbf {\bibinfo {volume} {41}},\ \bibinfo
  {pages} {429--452} (\bibinfo {year} {2012})}\BibitemShut {NoStop}%
\bibitem [{\citenamefont {Lindorff-Larsen}\ \emph {et~al.}(2011)\citenamefont
  {Lindorff-Larsen}, \citenamefont {Piana}, \citenamefont {Dror},\ and\
  \citenamefont {Shaw}}]{lindorff2011fast}%
  \BibitemOpen
  \bibfield  {author} {\bibinfo {author} {\bibfnamefont {K.}~\bibnamefont
  {Lindorff-Larsen}}, \bibinfo {author} {\bibfnamefont {S.}~\bibnamefont
  {Piana}}, \bibinfo {author} {\bibfnamefont {R.~O.}\ \bibnamefont {Dror}}, \
  and\ \bibinfo {author} {\bibfnamefont {D.~E.}\ \bibnamefont {Shaw}},\
  }\bibfield  {title} {\enquote {\bibinfo {title} {How fast-folding proteins
  fold},}\ }\href@noop {} {\bibfield  {journal} {\bibinfo  {journal} {Science}\
  }\textbf {\bibinfo {volume} {334}},\ \bibinfo {pages} {517--520} (\bibinfo
  {year} {2011})}\BibitemShut {NoStop}%
\bibitem [{\citenamefont {Baftizadeh}\ \emph {et~al.}(2012)\citenamefont
  {Baftizadeh}, \citenamefont {Biarnes}, \citenamefont {Pietrucci},
  \citenamefont {Affinito},\ and\ \citenamefont
  {Laio}}]{baftizadeh2012multidimensional}%
  \BibitemOpen
  \bibfield  {author} {\bibinfo {author} {\bibfnamefont {F.}~\bibnamefont
  {Baftizadeh}}, \bibinfo {author} {\bibfnamefont {X.}~\bibnamefont {Biarnes}},
  \bibinfo {author} {\bibfnamefont {F.}~\bibnamefont {Pietrucci}}, \bibinfo
  {author} {\bibfnamefont {F.}~\bibnamefont {Affinito}}, \ and\ \bibinfo
  {author} {\bibfnamefont {A.}~\bibnamefont {Laio}},\ }\bibfield  {title}
  {\enquote {\bibinfo {title} {Multidimensional view of amyloid fibril
  nucleation in atomistic detail},}\ }\href@noop {} {\bibfield  {journal}
  {\bibinfo  {journal} {J. Am. Chem. Soc.}\ }\textbf {\bibinfo {volume}
  {134}},\ \bibinfo {pages} {3886--3894} (\bibinfo {year} {2012})}\BibitemShut
  {NoStop}%
\bibitem [{\citenamefont {P{\'e}rez-Villa}, \citenamefont {Darvas},\ and\
  \citenamefont {Bussi}(2015)}]{perez2015atp}%
  \BibitemOpen
  \bibfield  {author} {\bibinfo {author} {\bibfnamefont {A.}~\bibnamefont
  {P{\'e}rez-Villa}}, \bibinfo {author} {\bibfnamefont {M.}~\bibnamefont
  {Darvas}}, \ and\ \bibinfo {author} {\bibfnamefont {G.}~\bibnamefont
  {Bussi}},\ }\bibfield  {title} {\enquote {\bibinfo {title} {{ATP} dependent
  {NS3} helicase interaction with {RNA}: insights from molecular
  simulations},}\ }\href@noop {} {\bibfield  {journal} {\bibinfo  {journal}
  {Nucleic Acids Res.}\ }\textbf {\bibinfo {volume} {43}},\ \bibinfo {pages}
  {8725--8734} (\bibinfo {year} {2015})}\BibitemShut {NoStop}%
\bibitem [{\citenamefont {Krepl}\ \emph {et~al.}(2015)\citenamefont {Krepl},
  \citenamefont {Havrila}, \citenamefont {Stadlbauer}, \citenamefont
  {Ban{\'a}{\v s}}, \citenamefont {Otyepka}, \citenamefont {Pasulka},
  \citenamefont {Stefl},\ and\ \citenamefont {{\v S}poner}}]{krepl2015can}%
  \BibitemOpen
  \bibfield  {author} {\bibinfo {author} {\bibfnamefont {M.}~\bibnamefont
  {Krepl}}, \bibinfo {author} {\bibfnamefont {M.}~\bibnamefont {Havrila}},
  \bibinfo {author} {\bibfnamefont {P.}~\bibnamefont {Stadlbauer}}, \bibinfo
  {author} {\bibfnamefont {P.}~\bibnamefont {Ban{\'a}{\v s}}}, \bibinfo
  {author} {\bibfnamefont {M.}~\bibnamefont {Otyepka}}, \bibinfo {author}
  {\bibfnamefont {J.}~\bibnamefont {Pasulka}}, \bibinfo {author} {\bibfnamefont
  {R.}~\bibnamefont {Stefl}}, \ and\ \bibinfo {author} {\bibfnamefont
  {J.}~\bibnamefont {{\v S}poner}},\ }\bibfield  {title} {\enquote {\bibinfo
  {title} {Can we execute stable microsecond-scale atomistic simulations of
  protein--{RNA} complexes?}}\ }\href@noop {} {\bibfield  {journal} {\bibinfo
  {journal} {J. Chem. Theory Comput.}\ }\textbf {\bibinfo {volume} {11}},\
  \bibinfo {pages} {1220--1243} (\bibinfo {year} {2015})}\BibitemShut {NoStop}%
\bibitem [{\citenamefont {Arkhipov}\ \emph {et~al.}(2013)\citenamefont
  {Arkhipov}, \citenamefont {Shan}, \citenamefont {Das}, \citenamefont
  {Endres}, \citenamefont {Eastwood}, \citenamefont {Wemmer}, \citenamefont
  {Kuriyan},\ and\ \citenamefont {Shaw}}]{arkhipov2013architecture}%
  \BibitemOpen
  \bibfield  {author} {\bibinfo {author} {\bibfnamefont {A.}~\bibnamefont
  {Arkhipov}}, \bibinfo {author} {\bibfnamefont {Y.}~\bibnamefont {Shan}},
  \bibinfo {author} {\bibfnamefont {R.}~\bibnamefont {Das}}, \bibinfo {author}
  {\bibfnamefont {N.~F.}\ \bibnamefont {Endres}}, \bibinfo {author}
  {\bibfnamefont {M.~P.}\ \bibnamefont {Eastwood}}, \bibinfo {author}
  {\bibfnamefont {D.~E.}\ \bibnamefont {Wemmer}}, \bibinfo {author}
  {\bibfnamefont {J.}~\bibnamefont {Kuriyan}}, \ and\ \bibinfo {author}
  {\bibfnamefont {D.~E.}\ \bibnamefont {Shaw}},\ }\bibfield  {title} {\enquote
  {\bibinfo {title} {Architecture and membrane interactions of the {EGF}
  receptor},}\ }\href@noop {} {\bibfield  {journal} {\bibinfo  {journal}
  {Cell}\ }\textbf {\bibinfo {volume} {152}},\ \bibinfo {pages} {557--569}
  (\bibinfo {year} {2013})}\BibitemShut {NoStop}%
\bibitem [{\citenamefont {Freddolino}\ \emph {et~al.}(2006)\citenamefont
  {Freddolino}, \citenamefont {Arkhipov}, \citenamefont {Larson}, \citenamefont
  {McPherson},\ and\ \citenamefont {Schulten}}]{freddolino2006molecular}%
  \BibitemOpen
  \bibfield  {author} {\bibinfo {author} {\bibfnamefont {P.~L.}\ \bibnamefont
  {Freddolino}}, \bibinfo {author} {\bibfnamefont {A.~S.}\ \bibnamefont
  {Arkhipov}}, \bibinfo {author} {\bibfnamefont {S.~B.}\ \bibnamefont
  {Larson}}, \bibinfo {author} {\bibfnamefont {A.}~\bibnamefont {McPherson}}, \
  and\ \bibinfo {author} {\bibfnamefont {K.}~\bibnamefont {Schulten}},\
  }\bibfield  {title} {\enquote {\bibinfo {title} {Molecular dynamics
  simulations of the complete satellite tobacco mosaic virus},}\ }\href@noop {}
  {\bibfield  {journal} {\bibinfo  {journal} {Structure}\ }\textbf {\bibinfo
  {volume} {14}},\ \bibinfo {pages} {437--449} (\bibinfo {year}
  {2006})}\BibitemShut {NoStop}%
\bibitem [{\citenamefont {Yu}\ \emph {et~al.}(2016)\citenamefont {Yu},
  \citenamefont {Mori}, \citenamefont {Ando}, \citenamefont {Harada},
  \citenamefont {Jung}, \citenamefont {Sugita},\ and\ \citenamefont
  {Feig}}]{yu2016biomolecular}%
  \BibitemOpen
  \bibfield  {author} {\bibinfo {author} {\bibfnamefont {I.}~\bibnamefont
  {Yu}}, \bibinfo {author} {\bibfnamefont {T.}~\bibnamefont {Mori}}, \bibinfo
  {author} {\bibfnamefont {T.}~\bibnamefont {Ando}}, \bibinfo {author}
  {\bibfnamefont {R.}~\bibnamefont {Harada}}, \bibinfo {author} {\bibfnamefont
  {J.}~\bibnamefont {Jung}}, \bibinfo {author} {\bibfnamefont {Y.}~\bibnamefont
  {Sugita}}, \ and\ \bibinfo {author} {\bibfnamefont {M.}~\bibnamefont
  {Feig}},\ }\bibfield  {title} {\enquote {\bibinfo {title} {Biomolecular
  interactions modulate macromolecular structure and dynamics in atomistic
  model of a bacterial cytoplasm},}\ }\href@noop {} {\bibfield  {journal}
  {\bibinfo  {journal} {Elife}\ }\textbf {\bibinfo {volume} {5}},\ \bibinfo
  {pages} {e19274} (\bibinfo {year} {2016})}\BibitemShut {NoStop}%
\bibitem [{\citenamefont {Singharoy}\ \emph {et~al.}(2019)\citenamefont
  {Singharoy}, \citenamefont {Maffeo}, \citenamefont {Delgado-Magnero},
  \citenamefont {Swainsbury}, \citenamefont {Sener}, \citenamefont
  {Kleinekath{\"o}fer}, \citenamefont {Vant}, \citenamefont {Nguyen},
  \citenamefont {Hitchcock}, \citenamefont {Isralewitz} \emph
  {et~al.}}]{singharoy2019atoms}%
  \BibitemOpen
  \bibfield  {author} {\bibinfo {author} {\bibfnamefont {A.}~\bibnamefont
  {Singharoy}}, \bibinfo {author} {\bibfnamefont {C.}~\bibnamefont {Maffeo}},
  \bibinfo {author} {\bibfnamefont {K.~H.}\ \bibnamefont {Delgado-Magnero}},
  \bibinfo {author} {\bibfnamefont {D.~J.}\ \bibnamefont {Swainsbury}},
  \bibinfo {author} {\bibfnamefont {M.}~\bibnamefont {Sener}}, \bibinfo
  {author} {\bibfnamefont {U.}~\bibnamefont {Kleinekath{\"o}fer}}, \bibinfo
  {author} {\bibfnamefont {J.~W.}\ \bibnamefont {Vant}}, \bibinfo {author}
  {\bibfnamefont {J.}~\bibnamefont {Nguyen}}, \bibinfo {author} {\bibfnamefont
  {A.}~\bibnamefont {Hitchcock}}, \bibinfo {author} {\bibfnamefont
  {B.}~\bibnamefont {Isralewitz}},  \emph {et~al.},\ }\bibfield  {title}
  {\enquote {\bibinfo {title} {Atoms to phenotypes: Molecular design principles
  of cellular energy metabolism},}\ }\href@noop {} {\bibfield  {journal}
  {\bibinfo  {journal} {Cell}\ }\textbf {\bibinfo {volume} {179}},\ \bibinfo
  {pages} {1098--1111} (\bibinfo {year} {2019})}\BibitemShut {NoStop}%
\bibitem [{\citenamefont {Shaw}\ \emph {et~al.}(2014)\citenamefont {Shaw},
  \citenamefont {Grossman}, \citenamefont {Bank}, \citenamefont {Batson},
  \citenamefont {Butts}, \citenamefont {Chao}, \citenamefont {Deneroff},
  \citenamefont {Dror}, \citenamefont {Even}, \citenamefont {Fenton} \emph
  {et~al.}}]{shaw2014anton}%
  \BibitemOpen
  \bibfield  {author} {\bibinfo {author} {\bibfnamefont {D.~E.}\ \bibnamefont
  {Shaw}}, \bibinfo {author} {\bibfnamefont {J.}~\bibnamefont {Grossman}},
  \bibinfo {author} {\bibfnamefont {J.~A.}\ \bibnamefont {Bank}}, \bibinfo
  {author} {\bibfnamefont {B.}~\bibnamefont {Batson}}, \bibinfo {author}
  {\bibfnamefont {J.~A.}\ \bibnamefont {Butts}}, \bibinfo {author}
  {\bibfnamefont {J.~C.}\ \bibnamefont {Chao}}, \bibinfo {author}
  {\bibfnamefont {M.~M.}\ \bibnamefont {Deneroff}}, \bibinfo {author}
  {\bibfnamefont {R.~O.}\ \bibnamefont {Dror}}, \bibinfo {author}
  {\bibfnamefont {A.}~\bibnamefont {Even}}, \bibinfo {author} {\bibfnamefont
  {C.~H.}\ \bibnamefont {Fenton}},  \emph {et~al.},\ }\bibfield  {title}
  {\enquote {\bibinfo {title} {Anton 2: raising the bar for performance and
  programmability in a special-purpose molecular dynamics supercomputer},}\
  }in\ \href@noop {} {\emph {\bibinfo {booktitle} {SC'14: Proceedings of the
  International Conference for High Performance Computing, Networking, Storage
  and Analysis}}}\ (\bibinfo {organization} {IEEE},\ \bibinfo {year} {2014})\
  pp.\ \bibinfo {pages} {41--53}\BibitemShut {NoStop}%
\bibitem [{\citenamefont {Case}\ \emph {et~al.}(2005)\citenamefont {Case},
  \citenamefont {Cheatham~III}, \citenamefont {Darden}, \citenamefont {Gohlke},
  \citenamefont {Luo}, \citenamefont {Merz~Jr}, \citenamefont {Onufriev},
  \citenamefont {Simmerling}, \citenamefont {Wang},\ and\ \citenamefont
  {Woods}}]{case2005amber}%
  \BibitemOpen
  \bibfield  {author} {\bibinfo {author} {\bibfnamefont {D.~A.}\ \bibnamefont
  {Case}}, \bibinfo {author} {\bibfnamefont {T.~E.}\ \bibnamefont
  {Cheatham~III}}, \bibinfo {author} {\bibfnamefont {T.}~\bibnamefont
  {Darden}}, \bibinfo {author} {\bibfnamefont {H.}~\bibnamefont {Gohlke}},
  \bibinfo {author} {\bibfnamefont {R.}~\bibnamefont {Luo}}, \bibinfo {author}
  {\bibfnamefont {K.~M.}\ \bibnamefont {Merz~Jr}}, \bibinfo {author}
  {\bibfnamefont {A.}~\bibnamefont {Onufriev}}, \bibinfo {author}
  {\bibfnamefont {C.}~\bibnamefont {Simmerling}}, \bibinfo {author}
  {\bibfnamefont {B.}~\bibnamefont {Wang}}, \ and\ \bibinfo {author}
  {\bibfnamefont {R.~J.}\ \bibnamefont {Woods}},\ }\bibfield  {title} {\enquote
  {\bibinfo {title} {The {A}mber biomolecular simulation programs},}\
  }\href@noop {} {\bibfield  {journal} {\bibinfo  {journal} {J. Comput. Chem.}\
  }\textbf {\bibinfo {volume} {26}},\ \bibinfo {pages} {1668--1688} (\bibinfo
  {year} {2005})}\BibitemShut {NoStop}%
\bibitem [{\citenamefont {Abraham}\ \emph {et~al.}(2015)\citenamefont
  {Abraham}, \citenamefont {Murtola}, \citenamefont {Schulz}, \citenamefont
  {P{\'a}ll}, \citenamefont {Smith}, \citenamefont {Hess},\ and\ \citenamefont
  {Lindahl}}]{abraham2015gromacs}%
  \BibitemOpen
  \bibfield  {author} {\bibinfo {author} {\bibfnamefont {M.~J.}\ \bibnamefont
  {Abraham}}, \bibinfo {author} {\bibfnamefont {T.}~\bibnamefont {Murtola}},
  \bibinfo {author} {\bibfnamefont {R.}~\bibnamefont {Schulz}}, \bibinfo
  {author} {\bibfnamefont {S.}~\bibnamefont {P{\'a}ll}}, \bibinfo {author}
  {\bibfnamefont {J.~C.}\ \bibnamefont {Smith}}, \bibinfo {author}
  {\bibfnamefont {B.}~\bibnamefont {Hess}}, \ and\ \bibinfo {author}
  {\bibfnamefont {E.}~\bibnamefont {Lindahl}},\ }\bibfield  {title} {\enquote
  {\bibinfo {title} {{GROMACS}: High performance molecular simulations through
  multi-level parallelism from laptops to supercomputers},}\ }\href@noop {}
  {\bibfield  {journal} {\bibinfo  {journal} {SoftwareX}\ }\textbf {\bibinfo
  {volume} {1}},\ \bibinfo {pages} {19--25} (\bibinfo {year}
  {2015})}\BibitemShut {NoStop}%
\bibitem [{\citenamefont {Valsson}, \citenamefont {Tiwary},\ and\ \citenamefont
  {Parrinello}(2016)}]{valsson2016enhancing}%
  \BibitemOpen
  \bibfield  {author} {\bibinfo {author} {\bibfnamefont {O.}~\bibnamefont
  {Valsson}}, \bibinfo {author} {\bibfnamefont {P.}~\bibnamefont {Tiwary}}, \
  and\ \bibinfo {author} {\bibfnamefont {M.}~\bibnamefont {Parrinello}},\
  }\bibfield  {title} {\enquote {\bibinfo {title} {Enhancing important
  fluctuations: Rare events and metadynamics from a conceptual viewpoint},}\
  }\href@noop {} {\bibfield  {journal} {\bibinfo  {journal} {Annu. Rev. Phys.
  Chem.}\ }\textbf {\bibinfo {volume} {67}},\ \bibinfo {pages} {159--184}
  (\bibinfo {year} {2016})}\BibitemShut {NoStop}%
\bibitem [{\citenamefont {Camilloni}\ and\ \citenamefont
  {Pietrucci}(2018)}]{camilloni2018advanced}%
  \BibitemOpen
  \bibfield  {author} {\bibinfo {author} {\bibfnamefont {C.}~\bibnamefont
  {Camilloni}}\ and\ \bibinfo {author} {\bibfnamefont {F.}~\bibnamefont
  {Pietrucci}},\ }\bibfield  {title} {\enquote {\bibinfo {title} {Advanced
  simulation techniques for the thermodynamic and kinetic characterization of
  biological systems},}\ }\href@noop {} {\bibfield  {journal} {\bibinfo
  {journal} {Adv. Phys. X}\ }\textbf {\bibinfo {volume} {3}},\ \bibinfo {pages}
  {1477531} (\bibinfo {year} {2018})}\BibitemShut {NoStop}%
\bibitem [{\citenamefont {Robustelli}, \citenamefont {Piana},\ and\
  \citenamefont {Shaw}(2018)}]{robustelli2018developing}%
  \BibitemOpen
  \bibfield  {author} {\bibinfo {author} {\bibfnamefont {P.}~\bibnamefont
  {Robustelli}}, \bibinfo {author} {\bibfnamefont {S.}~\bibnamefont {Piana}}, \
  and\ \bibinfo {author} {\bibfnamefont {D.~E.}\ \bibnamefont {Shaw}},\
  }\bibfield  {title} {\enquote {\bibinfo {title} {Developing a molecular
  dynamics force field for both folded and disordered protein states},}\
  }\href@noop {} {\bibfield  {journal} {\bibinfo  {journal} {Proc. Natl. Acad.
  Sci. U.S.A.}\ }\textbf {\bibinfo {volume} {115}},\ \bibinfo {pages}
  {E4758--E4766} (\bibinfo {year} {2018})}\BibitemShut {NoStop}%
\bibitem [{\citenamefont {Piana}\ \emph {et~al.}(2020)\citenamefont {Piana},
  \citenamefont {Robustelli}, \citenamefont {Tan}, \citenamefont {Chen},\ and\
  \citenamefont {Shaw}}]{piana2020development}%
  \BibitemOpen
  \bibfield  {author} {\bibinfo {author} {\bibfnamefont {S.}~\bibnamefont
  {Piana}}, \bibinfo {author} {\bibfnamefont {P.}~\bibnamefont {Robustelli}},
  \bibinfo {author} {\bibfnamefont {D.}~\bibnamefont {Tan}}, \bibinfo {author}
  {\bibfnamefont {S.}~\bibnamefont {Chen}}, \ and\ \bibinfo {author}
  {\bibfnamefont {D.~E.}\ \bibnamefont {Shaw}},\ }\bibfield  {title} {\enquote
  {\bibinfo {title} {Development of a force field for the simulation of
  single-chain proteins and protein-protein complexes},}\ }\href@noop {}
  {\bibfield  {journal} {\bibinfo  {journal} {J. Chem. Theory Comput.}\
  }\textbf {\bibinfo {volume} {16}},\ \bibinfo {pages} {2494--2507} (\bibinfo
  {year} {2020})}\BibitemShut {NoStop}%
\bibitem [{\citenamefont {{\v S}poner}\ \emph {et~al.}(2018)\citenamefont {{\v
  S}poner}, \citenamefont {Bussi}, \citenamefont {Krepl}, \citenamefont
  {Ban{\'a}{\v s}}, \citenamefont {Bottaro}, \citenamefont {Cunha},
  \citenamefont {Gil-Ley}, \citenamefont {Pinamonti}, \citenamefont {Poblete},
  \citenamefont {Jure\v{c}ka}, \citenamefont {Walter},\ and\ \citenamefont
  {Otyepka}}]{sponer2018rna}%
  \BibitemOpen
  \bibfield  {author} {\bibinfo {author} {\bibfnamefont {J.}~\bibnamefont {{\v
  S}poner}}, \bibinfo {author} {\bibfnamefont {G.}~\bibnamefont {Bussi}},
  \bibinfo {author} {\bibfnamefont {M.}~\bibnamefont {Krepl}}, \bibinfo
  {author} {\bibfnamefont {P.}~\bibnamefont {Ban{\'a}{\v s}}}, \bibinfo
  {author} {\bibfnamefont {S.}~\bibnamefont {Bottaro}}, \bibinfo {author}
  {\bibfnamefont {R.~A.}\ \bibnamefont {Cunha}}, \bibinfo {author}
  {\bibfnamefont {A.}~\bibnamefont {Gil-Ley}}, \bibinfo {author} {\bibfnamefont
  {G.}~\bibnamefont {Pinamonti}}, \bibinfo {author} {\bibfnamefont
  {S.}~\bibnamefont {Poblete}}, \bibinfo {author} {\bibfnamefont
  {P.}~\bibnamefont {Jure\v{c}ka}}, \bibinfo {author} {\bibfnamefont {N.~G.}\
  \bibnamefont {Walter}}, \ and\ \bibinfo {author} {\bibfnamefont
  {M.}~\bibnamefont {Otyepka}},\ }\bibfield  {title} {\enquote {\bibinfo
  {title} {{RNA} structural dynamics as captured by molecular simulations: a
  comprehensive overview},}\ }\href@noop {} {\bibfield  {journal} {\bibinfo
  {journal} {Chem. Rev.}\ }\textbf {\bibinfo {volume} {118}},\ \bibinfo {pages}
  {4177--4338} (\bibinfo {year} {2018})}\BibitemShut {NoStop}%
\bibitem [{\citenamefont {Capelli}\ \emph {et~al.}(2020)\citenamefont
  {Capelli}, \citenamefont {Lyu}, \citenamefont {Bolnykh}, \citenamefont
  {Meloni}, \citenamefont {Olsen}, \citenamefont {Rothlisberger}, \citenamefont
  {Parrinello},\ and\ \citenamefont {Carloni}}]{capelli2020accuracy}%
  \BibitemOpen
  \bibfield  {author} {\bibinfo {author} {\bibfnamefont {R.}~\bibnamefont
  {Capelli}}, \bibinfo {author} {\bibfnamefont {W.}~\bibnamefont {Lyu}},
  \bibinfo {author} {\bibfnamefont {V.}~\bibnamefont {Bolnykh}}, \bibinfo
  {author} {\bibfnamefont {S.}~\bibnamefont {Meloni}}, \bibinfo {author}
  {\bibfnamefont {J.~M.~H.}\ \bibnamefont {Olsen}}, \bibinfo {author}
  {\bibfnamefont {U.}~\bibnamefont {Rothlisberger}}, \bibinfo {author}
  {\bibfnamefont {M.}~\bibnamefont {Parrinello}}, \ and\ \bibinfo {author}
  {\bibfnamefont {P.}~\bibnamefont {Carloni}},\ }\bibfield  {title} {\enquote
  {\bibinfo {title} {On the accuracy of molecular simulation-based predictions
  of koff values: a metadynamics study},}\ }\href@noop {} {\bibfield  {journal}
  {\bibinfo  {journal} {bioRxiv preprint, doi:10.1101/2020.03.30.015396}\ }
  (\bibinfo {year} {2020})}\BibitemShut {NoStop}%
\bibitem [{\citenamefont {Cornell}\ \emph {et~al.}(1995)\citenamefont
  {Cornell}, \citenamefont {Cieplak}, \citenamefont {Bayly}, \citenamefont
  {Gould}, \citenamefont {Merz}, \citenamefont {Ferguson}, \citenamefont
  {Spellmeyer}, \citenamefont {Fox}, \citenamefont {Caldwell},\ and\
  \citenamefont {Kollman}}]{cornell1995amber}%
  \BibitemOpen
  \bibfield  {author} {\bibinfo {author} {\bibfnamefont {W.~D.}\ \bibnamefont
  {Cornell}}, \bibinfo {author} {\bibfnamefont {P.}~\bibnamefont {Cieplak}},
  \bibinfo {author} {\bibfnamefont {C.~I.}\ \bibnamefont {Bayly}}, \bibinfo
  {author} {\bibfnamefont {I.~R.}\ \bibnamefont {Gould}}, \bibinfo {author}
  {\bibfnamefont {K.~M.}\ \bibnamefont {Merz}}, \bibinfo {author}
  {\bibfnamefont {D.~M.}\ \bibnamefont {Ferguson}}, \bibinfo {author}
  {\bibfnamefont {D.~C.}\ \bibnamefont {Spellmeyer}}, \bibinfo {author}
  {\bibfnamefont {T.}~\bibnamefont {Fox}}, \bibinfo {author} {\bibfnamefont
  {J.~W.}\ \bibnamefont {Caldwell}}, \ and\ \bibinfo {author} {\bibfnamefont
  {P.~A.}\ \bibnamefont {Kollman}},\ }\bibfield  {title} {\enquote {\bibinfo
  {title} {A second generation force field for the simulation of proteins,
  nucleic acids, and organic molecules},}\ }\href@noop {} {\bibfield  {journal}
  {\bibinfo  {journal} {J. Am. Chem. Soc.}\ }\textbf {\bibinfo {volume}
  {117}},\ \bibinfo {pages} {5179--5197} (\bibinfo {year} {1995})}\BibitemShut
  {NoStop}%
\bibitem [{\citenamefont {MacKerell}, \citenamefont {Wiorkiewicz-Kuczera},\
  and\ \citenamefont {Karplus}(1995)}]{mackerell1995charmm}%
  \BibitemOpen
  \bibfield  {author} {\bibinfo {author} {\bibfnamefont {A.~D.}\ \bibnamefont
  {MacKerell}}, \bibinfo {author} {\bibfnamefont {J.}~\bibnamefont
  {Wiorkiewicz-Kuczera}}, \ and\ \bibinfo {author} {\bibfnamefont
  {M.}~\bibnamefont {Karplus}},\ }\bibfield  {title} {\enquote {\bibinfo
  {title} {An all-atom empirical energy function for the simulation of nucleic
  acids},}\ }\href@noop {} {\bibfield  {journal} {\bibinfo  {journal} {J. Am.
  Chem. Soc.}\ }\textbf {\bibinfo {volume} {117}},\ \bibinfo {pages}
  {11946--11975} (\bibinfo {year} {1995})}\BibitemShut {NoStop}%
\bibitem [{\citenamefont {Jorgensen}\ and\ \citenamefont
  {Tirado-Rives}(1988)}]{jorgensen1988opls}%
  \BibitemOpen
  \bibfield  {author} {\bibinfo {author} {\bibfnamefont {W.~L.}\ \bibnamefont
  {Jorgensen}}\ and\ \bibinfo {author} {\bibfnamefont {J.}~\bibnamefont
  {Tirado-Rives}},\ }\bibfield  {title} {\enquote {\bibinfo {title} {The {OPLS}
  [optimized potentials for liquid simulations] potential functions for
  proteins, energy minimizations for crystals of cyclic peptides and
  crambin},}\ }\href@noop {} {\bibfield  {journal} {\bibinfo  {journal} {J. Am.
  Chem. Soc.}\ }\textbf {\bibinfo {volume} {110}},\ \bibinfo {pages}
  {1657--1666} (\bibinfo {year} {1988})}\BibitemShut {NoStop}%
\bibitem [{\citenamefont {Oostenbrink}\ \emph {et~al.}(2004)\citenamefont
  {Oostenbrink}, \citenamefont {Villa}, \citenamefont {Mark},\ and\
  \citenamefont {Van~Gunsteren}}]{oostenbrink2004gromos}%
  \BibitemOpen
  \bibfield  {author} {\bibinfo {author} {\bibfnamefont {C.}~\bibnamefont
  {Oostenbrink}}, \bibinfo {author} {\bibfnamefont {A.}~\bibnamefont {Villa}},
  \bibinfo {author} {\bibfnamefont {A.~E.}\ \bibnamefont {Mark}}, \ and\
  \bibinfo {author} {\bibfnamefont {W.~F.}\ \bibnamefont {Van~Gunsteren}},\
  }\bibfield  {title} {\enquote {\bibinfo {title} {A biomolecular force field
  based on the free enthalpy of hydration and solvation: The {GROMOS}
  force-field parameter sets {53A5} and {53A6}},}\ }\href@noop {} {\bibfield
  {journal} {\bibinfo  {journal} {J. Comput. Chem.}\ }\textbf {\bibinfo
  {volume} {25}},\ \bibinfo {pages} {1656--1676} (\bibinfo {year}
  {2004})}\BibitemShut {NoStop}%
\bibitem [{\citenamefont {Shi}\ \emph {et~al.}(2013)\citenamefont {Shi},
  \citenamefont {Xia}, \citenamefont {Zhang}, \citenamefont {Best},
  \citenamefont {Wu}, \citenamefont {Ponder},\ and\ \citenamefont
  {Ren}}]{shi2013polarizable}%
  \BibitemOpen
  \bibfield  {author} {\bibinfo {author} {\bibfnamefont {Y.}~\bibnamefont
  {Shi}}, \bibinfo {author} {\bibfnamefont {Z.}~\bibnamefont {Xia}}, \bibinfo
  {author} {\bibfnamefont {J.}~\bibnamefont {Zhang}}, \bibinfo {author}
  {\bibfnamefont {R.}~\bibnamefont {Best}}, \bibinfo {author} {\bibfnamefont
  {C.}~\bibnamefont {Wu}}, \bibinfo {author} {\bibfnamefont {J.~W.}\
  \bibnamefont {Ponder}}, \ and\ \bibinfo {author} {\bibfnamefont
  {P.}~\bibnamefont {Ren}},\ }\bibfield  {title} {\enquote {\bibinfo {title}
  {Polarizable atomic multipole-based {AMOEBA} force field for proteins},}\
  }\href@noop {} {\bibfield  {journal} {\bibinfo  {journal} {J. Chem. Theory
  Comput.}\ }\textbf {\bibinfo {volume} {9}},\ \bibinfo {pages} {4046--4063}
  (\bibinfo {year} {2013})}\BibitemShut {NoStop}%
\bibitem [{\citenamefont {Patel}, \citenamefont {Mackerell~Jr},\ and\
  \citenamefont {Brooks~III}(2004)}]{patel2004charmm}%
  \BibitemOpen
  \bibfield  {author} {\bibinfo {author} {\bibfnamefont {S.}~\bibnamefont
  {Patel}}, \bibinfo {author} {\bibfnamefont {A.~D.}\ \bibnamefont
  {Mackerell~Jr}}, \ and\ \bibinfo {author} {\bibfnamefont {C.~L.}\
  \bibnamefont {Brooks~III}},\ }\bibfield  {title} {\enquote {\bibinfo {title}
  {{CHARMM} fluctuating charge force field for proteins: {II} protein/solvent
  properties from molecular dynamics simulations using a nonadditive
  electrostatic model},}\ }\href@noop {} {\bibfield  {journal} {\bibinfo
  {journal} {J. Comput. Chem.}\ }\textbf {\bibinfo {volume} {25}},\ \bibinfo
  {pages} {1504--1514} (\bibinfo {year} {2004})}\BibitemShut {NoStop}%
\bibitem [{\citenamefont {Senftle}\ \emph {et~al.}(2016)\citenamefont
  {Senftle}, \citenamefont {Hong}, \citenamefont {Islam}, \citenamefont
  {Kylasa}, \citenamefont {Zheng}, \citenamefont {Shin}, \citenamefont
  {Junkermeier}, \citenamefont {Engel-Herbert}, \citenamefont {Janik},
  \citenamefont {Aktulga} \emph {et~al.}}]{senftle2016reaxff}%
  \BibitemOpen
  \bibfield  {author} {\bibinfo {author} {\bibfnamefont {T.~P.}\ \bibnamefont
  {Senftle}}, \bibinfo {author} {\bibfnamefont {S.}~\bibnamefont {Hong}},
  \bibinfo {author} {\bibfnamefont {M.~M.}\ \bibnamefont {Islam}}, \bibinfo
  {author} {\bibfnamefont {S.~B.}\ \bibnamefont {Kylasa}}, \bibinfo {author}
  {\bibfnamefont {Y.}~\bibnamefont {Zheng}}, \bibinfo {author} {\bibfnamefont
  {Y.~K.}\ \bibnamefont {Shin}}, \bibinfo {author} {\bibfnamefont
  {C.}~\bibnamefont {Junkermeier}}, \bibinfo {author} {\bibfnamefont
  {R.}~\bibnamefont {Engel-Herbert}}, \bibinfo {author} {\bibfnamefont {M.~J.}\
  \bibnamefont {Janik}}, \bibinfo {author} {\bibfnamefont {H.~M.}\ \bibnamefont
  {Aktulga}},  \emph {et~al.},\ }\bibfield  {title} {\enquote {\bibinfo {title}
  {The {ReaxFF} reactive force-field: development, applications and future
  directions},}\ }\href@noop {} {\bibfield  {journal} {\bibinfo  {journal} {NPJ
  Comput. Mater.}\ }\textbf {\bibinfo {volume} {2}},\ \bibinfo {pages} {1--14}
  (\bibinfo {year} {2016})}\BibitemShut {NoStop}%
\bibitem [{\citenamefont {Seifert}\ and\ \citenamefont
  {Joswig}(2012)}]{seifert2012density}%
  \BibitemOpen
  \bibfield  {author} {\bibinfo {author} {\bibfnamefont {G.}~\bibnamefont
  {Seifert}}\ and\ \bibinfo {author} {\bibfnamefont {J.-O.}\ \bibnamefont
  {Joswig}},\ }\bibfield  {title} {\enquote {\bibinfo {title}
  {Density-functional tight binding -- an approximate density-functional theory
  method},}\ }\href@noop {} {\bibfield  {journal} {\bibinfo  {journal} {Wiley
  Interdiscip. Rev. Comput. Mol. Sci.}\ }\textbf {\bibinfo {volume} {2}},\
  \bibinfo {pages} {456--465} (\bibinfo {year} {2012})}\BibitemShut {NoStop}%
\bibitem [{\citenamefont {V{\'a}rnai}\ and\ \citenamefont
  {Zakrzewska}(2004)}]{varnai2004dna}%
  \BibitemOpen
  \bibfield  {author} {\bibinfo {author} {\bibfnamefont {P.}~\bibnamefont
  {V{\'a}rnai}}\ and\ \bibinfo {author} {\bibfnamefont {K.}~\bibnamefont
  {Zakrzewska}},\ }\bibfield  {title} {\enquote {\bibinfo {title} {{DNA} and
  its counterions: a molecular dynamics study},}\ }\href@noop {} {\bibfield
  {journal} {\bibinfo  {journal} {Nucleic Acids Res.}\ }\textbf {\bibinfo
  {volume} {32}},\ \bibinfo {pages} {4269--4280} (\bibinfo {year}
  {2004})}\BibitemShut {NoStop}%
\bibitem [{\citenamefont {P{\'e}rez}\ \emph {et~al.}(2007)\citenamefont
  {P{\'e}rez}, \citenamefont {March{\'a}n}, \citenamefont {Svozil},
  \citenamefont {{\v S}poner}, \citenamefont {Cheatham~III}, \citenamefont
  {Laughton},\ and\ \citenamefont {Orozco}}]{perez2007refinement}%
  \BibitemOpen
  \bibfield  {author} {\bibinfo {author} {\bibfnamefont {A.}~\bibnamefont
  {P{\'e}rez}}, \bibinfo {author} {\bibfnamefont {I.}~\bibnamefont
  {March{\'a}n}}, \bibinfo {author} {\bibfnamefont {D.}~\bibnamefont {Svozil}},
  \bibinfo {author} {\bibfnamefont {J.}~\bibnamefont {{\v S}poner}}, \bibinfo
  {author} {\bibfnamefont {T.~E.}\ \bibnamefont {Cheatham~III}}, \bibinfo
  {author} {\bibfnamefont {C.~A.}\ \bibnamefont {Laughton}}, \ and\ \bibinfo
  {author} {\bibfnamefont {M.}~\bibnamefont {Orozco}},\ }\bibfield  {title}
  {\enquote {\bibinfo {title} {Refinement of the {AMBER} force field for
  nucleic acids: improving the description of $\alpha$/$\gamma$ conformers},}\
  }\href@noop {} {\bibfield  {journal} {\bibinfo  {journal} {Biophys. J.}\
  }\textbf {\bibinfo {volume} {92}},\ \bibinfo {pages} {3817--3829} (\bibinfo
  {year} {2007})}\BibitemShut {NoStop}%
\bibitem [{\citenamefont {Zgarbov\'a}\ \emph {et~al.}(2011)\citenamefont
  {Zgarbov\'a}, \citenamefont {Otyepka}, \citenamefont {\v{S}poner},
  \citenamefont {Ml\'{a}dek}, \citenamefont {Ban\'{a}\v{s}}, \citenamefont
  {Cheatham},\ and\ \citenamefont {Jure\v{c}ka}}]{zgarbova2011refinement}%
  \BibitemOpen
  \bibfield  {author} {\bibinfo {author} {\bibfnamefont {M.}~\bibnamefont
  {Zgarbov\'a}}, \bibinfo {author} {\bibfnamefont {M.}~\bibnamefont {Otyepka}},
  \bibinfo {author} {\bibfnamefont {J.}~\bibnamefont {\v{S}poner}}, \bibinfo
  {author} {\bibfnamefont {A.}~\bibnamefont {Ml\'{a}dek}}, \bibinfo {author}
  {\bibfnamefont {P.}~\bibnamefont {Ban\'{a}\v{s}}}, \bibinfo {author}
  {\bibfnamefont {T.~E.}\ \bibnamefont {Cheatham}}, \ and\ \bibinfo {author}
  {\bibfnamefont {P.}~\bibnamefont {Jure\v{c}ka}},\ }\bibfield  {title}
  {\enquote {\bibinfo {title} {Refinement of the {C}ornell et al. nucleic acids
  force field based on reference quantum chemical calculations of glycosidic
  torsion profiles},}\ }\href {\doibase 10.1021/ct200162x} {\bibfield
  {journal} {\bibinfo  {journal} {J. Chem. Theory Comput.}\ }\textbf {\bibinfo
  {volume} {7}},\ \bibinfo {pages} {2886--2902} (\bibinfo {year}
  {2011})}\BibitemShut {NoStop}%
\bibitem [{\citenamefont {Mackerell~Jr}, \citenamefont {Feig},\ and\
  \citenamefont {Brooks~III}(2004)}]{mackerell2004extending}%
  \BibitemOpen
  \bibfield  {author} {\bibinfo {author} {\bibfnamefont {A.~D.}\ \bibnamefont
  {Mackerell~Jr}}, \bibinfo {author} {\bibfnamefont {M.}~\bibnamefont {Feig}},
  \ and\ \bibinfo {author} {\bibfnamefont {C.~L.}\ \bibnamefont {Brooks~III}},\
  }\bibfield  {title} {\enquote {\bibinfo {title} {Extending the treatment of
  backbone energetics in protein force fields: limitations of gas-phase quantum
  mechanics in reproducing protein conformational distributions in molecular
  dynamics simulations},}\ }\href@noop {} {\bibfield  {journal} {\bibinfo
  {journal} {J. Comput. Chem.}\ }\textbf {\bibinfo {volume} {25}},\ \bibinfo
  {pages} {1400--1415} (\bibinfo {year} {2004})}\BibitemShut {NoStop}%
\bibitem [{\citenamefont {Best}\ and\ \citenamefont
  {Hummer}(2009)}]{best2009optimized}%
  \BibitemOpen
  \bibfield  {author} {\bibinfo {author} {\bibfnamefont {R.~B.}\ \bibnamefont
  {Best}}\ and\ \bibinfo {author} {\bibfnamefont {G.}~\bibnamefont {Hummer}},\
  }\bibfield  {title} {\enquote {\bibinfo {title} {Optimized molecular dynamics
  force fields applied to the helix-coil transition of polypeptides},}\
  }\href@noop {} {\bibfield  {journal} {\bibinfo  {journal} {J. Phys. Chem. B}\
  }\textbf {\bibinfo {volume} {113}},\ \bibinfo {pages} {9004--9015} (\bibinfo
  {year} {2009})}\BibitemShut {NoStop}%
\bibitem [{\citenamefont {Piana}, \citenamefont {Lindorff-Larsen},\ and\
  \citenamefont {Shaw}(2011)}]{piana2011robust}%
  \BibitemOpen
  \bibfield  {author} {\bibinfo {author} {\bibfnamefont {S.}~\bibnamefont
  {Piana}}, \bibinfo {author} {\bibfnamefont {K.}~\bibnamefont
  {Lindorff-Larsen}}, \ and\ \bibinfo {author} {\bibfnamefont {D.~E.}\
  \bibnamefont {Shaw}},\ }\bibfield  {title} {\enquote {\bibinfo {title} {How
  robust are protein folding simulations with respect to force field
  parameterization?}}\ }\href@noop {} {\bibfield  {journal} {\bibinfo
  {journal} {Biophys. J.}\ }\textbf {\bibinfo {volume} {100}},\ \bibinfo
  {pages} {L47--L49} (\bibinfo {year} {2011})}\BibitemShut {NoStop}%
\bibitem [{\citenamefont {Best}\ \emph {et~al.}(2012)\citenamefont {Best},
  \citenamefont {Zhu}, \citenamefont {Shim}, \citenamefont {Lopes},
  \citenamefont {Mittal}, \citenamefont {Feig},\ and\ \citenamefont
  {MacKerell~Jr}}]{best2012optimization}%
  \BibitemOpen
  \bibfield  {author} {\bibinfo {author} {\bibfnamefont {R.~B.}\ \bibnamefont
  {Best}}, \bibinfo {author} {\bibfnamefont {X.}~\bibnamefont {Zhu}}, \bibinfo
  {author} {\bibfnamefont {J.}~\bibnamefont {Shim}}, \bibinfo {author}
  {\bibfnamefont {P.~E.}\ \bibnamefont {Lopes}}, \bibinfo {author}
  {\bibfnamefont {J.}~\bibnamefont {Mittal}}, \bibinfo {author} {\bibfnamefont
  {M.}~\bibnamefont {Feig}}, \ and\ \bibinfo {author} {\bibfnamefont {A.~D.}\
  \bibnamefont {MacKerell~Jr}},\ }\bibfield  {title} {\enquote {\bibinfo
  {title} {Optimization of the additive {CHARMM} all-atom protein force field
  targeting improved sampling of the backbone $\phi$, $\psi$ and side-chain
  $\chi$1 and $\chi$2 dihedral angles},}\ }\href@noop {} {\bibfield  {journal}
  {\bibinfo  {journal} {J. Chem Theory Comput.}\ }\textbf {\bibinfo {volume}
  {8}},\ \bibinfo {pages} {3257--3273} (\bibinfo {year} {2012})}\BibitemShut
  {NoStop}%
\bibitem [{\citenamefont {Norgaard}, \citenamefont {Ferkinghoff-Borg},\ and\
  \citenamefont {Lindorff-Larsen}(2008)}]{norgaard2008experimental}%
  \BibitemOpen
  \bibfield  {author} {\bibinfo {author} {\bibfnamefont {A.~B.}\ \bibnamefont
  {Norgaard}}, \bibinfo {author} {\bibfnamefont {J.}~\bibnamefont
  {Ferkinghoff-Borg}}, \ and\ \bibinfo {author} {\bibfnamefont
  {K.}~\bibnamefont {Lindorff-Larsen}},\ }\bibfield  {title} {\enquote
  {\bibinfo {title} {Experimental parameterization of an energy function for
  the simulation of unfolded proteins},}\ }\href@noop {} {\bibfield  {journal}
  {\bibinfo  {journal} {Biophys. J.}\ }\textbf {\bibinfo {volume} {94}},\
  \bibinfo {pages} {182--192} (\bibinfo {year} {2008})}\BibitemShut {NoStop}%
\bibitem [{\citenamefont {Bottaro}, \citenamefont {Lindorff-Larsen},\ and\
  \citenamefont {Best}(2013)}]{bottaro2013variational}%
  \BibitemOpen
  \bibfield  {author} {\bibinfo {author} {\bibfnamefont {S.}~\bibnamefont
  {Bottaro}}, \bibinfo {author} {\bibfnamefont {K.}~\bibnamefont
  {Lindorff-Larsen}}, \ and\ \bibinfo {author} {\bibfnamefont {R.~B.}\
  \bibnamefont {Best}},\ }\bibfield  {title} {\enquote {\bibinfo {title}
  {Variational optimization of an all-atom implicit solvent force field to
  match explicit solvent simulation data},}\ }\href@noop {} {\bibfield
  {journal} {\bibinfo  {journal} {J. Chem Theory Comput.}\ }\textbf {\bibinfo
  {volume} {9}},\ \bibinfo {pages} {5641--5652} (\bibinfo {year}
  {2013})}\BibitemShut {NoStop}%
\bibitem [{\citenamefont {Li}\ and\ \citenamefont
  {Br{\"u}schweiler}(2011)}]{li2011iterative}%
  \BibitemOpen
  \bibfield  {author} {\bibinfo {author} {\bibfnamefont {D.-W.}\ \bibnamefont
  {Li}}\ and\ \bibinfo {author} {\bibfnamefont {R.}~\bibnamefont
  {Br{\"u}schweiler}},\ }\bibfield  {title} {\enquote {\bibinfo {title}
  {Iterative optimization of molecular mechanics force fields from {NMR} data
  of full-length proteins},}\ }\href@noop {} {\bibfield  {journal} {\bibinfo
  {journal} {J. Chem. Theory Comput.}\ }\textbf {\bibinfo {volume} {7}},\
  \bibinfo {pages} {1773--1782} (\bibinfo {year} {2011})}\BibitemShut {NoStop}%
\bibitem [{\citenamefont {Wang}, \citenamefont {Chen},\ and\ \citenamefont
  {Van~Voorhis}(2012)}]{wang2012systematic}%
  \BibitemOpen
  \bibfield  {author} {\bibinfo {author} {\bibfnamefont {L.-P.}\ \bibnamefont
  {Wang}}, \bibinfo {author} {\bibfnamefont {J.}~\bibnamefont {Chen}}, \ and\
  \bibinfo {author} {\bibfnamefont {T.}~\bibnamefont {Van~Voorhis}},\
  }\bibfield  {title} {\enquote {\bibinfo {title} {Systematic parametrization
  of polarizable force fields from quantum chemistry data},}\ }\href@noop {}
  {\bibfield  {journal} {\bibinfo  {journal} {J. Chem. Theory Comput.}\
  }\textbf {\bibinfo {volume} {9}},\ \bibinfo {pages} {452--460} (\bibinfo
  {year} {2012})}\BibitemShut {NoStop}%
\bibitem [{\citenamefont {Wang}, \citenamefont {Martinez},\ and\ \citenamefont
  {Pande}(2014)}]{wang2014building}%
  \BibitemOpen
  \bibfield  {author} {\bibinfo {author} {\bibfnamefont {L.-P.}\ \bibnamefont
  {Wang}}, \bibinfo {author} {\bibfnamefont {T.~J.}\ \bibnamefont {Martinez}},
  \ and\ \bibinfo {author} {\bibfnamefont {V.~S.}\ \bibnamefont {Pande}},\
  }\bibfield  {title} {\enquote {\bibinfo {title} {Building force fields: an
  automatic, systematic, and reproducible approach},}\ }\href@noop {}
  {\bibfield  {journal} {\bibinfo  {journal} {J. Phys. Chem. Lett.}\ }\textbf
  {\bibinfo {volume} {5}},\ \bibinfo {pages} {1885--1891} (\bibinfo {year}
  {2014})}\BibitemShut {NoStop}%
\bibitem [{\citenamefont {Cesari}\ \emph {et~al.}(2019)\citenamefont {Cesari},
  \citenamefont {Bottaro}, \citenamefont {Lindorff-Larsen}, \citenamefont
  {Ban{\'a}{\v s}}, \citenamefont {{\v S}poner},\ and\ \citenamefont
  {Bussi}}]{cesari2019fitting}%
  \BibitemOpen
  \bibfield  {author} {\bibinfo {author} {\bibfnamefont {A.}~\bibnamefont
  {Cesari}}, \bibinfo {author} {\bibfnamefont {S.}~\bibnamefont {Bottaro}},
  \bibinfo {author} {\bibfnamefont {K.}~\bibnamefont {Lindorff-Larsen}},
  \bibinfo {author} {\bibfnamefont {P.}~\bibnamefont {Ban{\'a}{\v s}}},
  \bibinfo {author} {\bibfnamefont {J.}~\bibnamefont {{\v S}poner}}, \ and\
  \bibinfo {author} {\bibfnamefont {G.}~\bibnamefont {Bussi}},\ }\bibfield
  {title} {\enquote {\bibinfo {title} {Fitting corrections to an {RNA} force
  field using experimental data},}\ }\href@noop {} {\bibfield  {journal}
  {\bibinfo  {journal} {J. Chem. Theory Comput.}\ }\textbf {\bibinfo {volume}
  {15}},\ \bibinfo {pages} {3425--3431} (\bibinfo {year} {2019})}\BibitemShut
  {NoStop}%
\bibitem [{\citenamefont {Bonomi}\ \emph {et~al.}(2017)\citenamefont {Bonomi},
  \citenamefont {Heller}, \citenamefont {Camilloni},\ and\ \citenamefont
  {Vendruscolo}}]{bonomi2017principles}%
  \BibitemOpen
  \bibfield  {author} {\bibinfo {author} {\bibfnamefont {M.}~\bibnamefont
  {Bonomi}}, \bibinfo {author} {\bibfnamefont {G.~T.}\ \bibnamefont {Heller}},
  \bibinfo {author} {\bibfnamefont {C.}~\bibnamefont {Camilloni}}, \ and\
  \bibinfo {author} {\bibfnamefont {M.}~\bibnamefont {Vendruscolo}},\
  }\bibfield  {title} {\enquote {\bibinfo {title} {Principles of protein
  structural ensemble determination},}\ }\href@noop {} {\bibfield  {journal}
  {\bibinfo  {journal} {Curr. Opin. Struct. Biol.}\ }\textbf {\bibinfo {volume}
  {42}},\ \bibinfo {pages} {106--116} (\bibinfo {year} {2017})}\BibitemShut
  {NoStop}%
\bibitem [{\citenamefont {Hummer}\ and\ \citenamefont
  {K{\"o}finger}(2015)}]{hummer2015bayesian}%
  \BibitemOpen
  \bibfield  {author} {\bibinfo {author} {\bibfnamefont {G.}~\bibnamefont
  {Hummer}}\ and\ \bibinfo {author} {\bibfnamefont {J.}~\bibnamefont
  {K{\"o}finger}},\ }\bibfield  {title} {\enquote {\bibinfo {title} {Bayesian
  ensemble refinement by replica simulations and reweighting},}\ }\href@noop {}
  {\bibfield  {journal} {\bibinfo  {journal} {J. Chem. Phys.}\ }\textbf
  {\bibinfo {volume} {143}},\ \bibinfo {pages} {12B634\_1} (\bibinfo {year}
  {2015})}\BibitemShut {NoStop}%
\bibitem [{\citenamefont {Bonomi}\ \emph {et~al.}(2016)\citenamefont {Bonomi},
  \citenamefont {Camilloni}, \citenamefont {Cavalli},\ and\ \citenamefont
  {Vendruscolo}}]{bonomi2016metainference}%
  \BibitemOpen
  \bibfield  {author} {\bibinfo {author} {\bibfnamefont {M.}~\bibnamefont
  {Bonomi}}, \bibinfo {author} {\bibfnamefont {C.}~\bibnamefont {Camilloni}},
  \bibinfo {author} {\bibfnamefont {A.}~\bibnamefont {Cavalli}}, \ and\
  \bibinfo {author} {\bibfnamefont {M.}~\bibnamefont {Vendruscolo}},\
  }\bibfield  {title} {\enquote {\bibinfo {title} {Metainference: A {B}ayesian
  inference method for heterogeneous systems},}\ }\href@noop {} {\bibfield
  {journal} {\bibinfo  {journal} {Sci. Adv.}\ }\textbf {\bibinfo {volume}
  {2}},\ \bibinfo {pages} {e1501177} (\bibinfo {year} {2016})}\BibitemShut
  {NoStop}%
\bibitem [{\citenamefont {Pitera}\ and\ \citenamefont
  {Chodera}(2012)}]{pitera2012use}%
  \BibitemOpen
  \bibfield  {author} {\bibinfo {author} {\bibfnamefont {J.~W.}\ \bibnamefont
  {Pitera}}\ and\ \bibinfo {author} {\bibfnamefont {J.~D.}\ \bibnamefont
  {Chodera}},\ }\bibfield  {title} {\enquote {\bibinfo {title} {On the use of
  experimental observations to bias simulated ensembles},}\ }\href@noop {}
  {\bibfield  {journal} {\bibinfo  {journal} {J. Chem. Theory Comput.}\
  }\textbf {\bibinfo {volume} {8}},\ \bibinfo {pages} {3445--3451} (\bibinfo
  {year} {2012})}\BibitemShut {NoStop}%
\bibitem [{\citenamefont {Cesari}, \citenamefont {Rei{\ss}er},\ and\
  \citenamefont {Bussi}(2018)}]{cesari2018using}%
  \BibitemOpen
  \bibfield  {author} {\bibinfo {author} {\bibfnamefont {A.}~\bibnamefont
  {Cesari}}, \bibinfo {author} {\bibfnamefont {S.}~\bibnamefont {Rei{\ss}er}},
  \ and\ \bibinfo {author} {\bibfnamefont {G.}~\bibnamefont {Bussi}},\
  }\bibfield  {title} {\enquote {\bibinfo {title} {Using the maximum entropy
  principle to combine simulations and solution experiments},}\ }\href@noop {}
  {\bibfield  {journal} {\bibinfo  {journal} {Computation}\ }\textbf {\bibinfo
  {volume} {6}},\ \bibinfo {pages} {15} (\bibinfo {year} {2018})}\BibitemShut
  {NoStop}%
\bibitem [{\citenamefont {Bottaro}\ \emph {et~al.}(2018)\citenamefont
  {Bottaro}, \citenamefont {Bussi}, \citenamefont {Kennedy}, \citenamefont
  {Turner},\ and\ \citenamefont {Lindorff-Larsen}}]{bottaro2018conformational}%
  \BibitemOpen
  \bibfield  {author} {\bibinfo {author} {\bibfnamefont {S.}~\bibnamefont
  {Bottaro}}, \bibinfo {author} {\bibfnamefont {G.}~\bibnamefont {Bussi}},
  \bibinfo {author} {\bibfnamefont {S.~D.}\ \bibnamefont {Kennedy}}, \bibinfo
  {author} {\bibfnamefont {D.~H.}\ \bibnamefont {Turner}}, \ and\ \bibinfo
  {author} {\bibfnamefont {K.}~\bibnamefont {Lindorff-Larsen}},\ }\bibfield
  {title} {\enquote {\bibinfo {title} {Conformational ensembles of {RNA}
  oligonucleotides from integrating {NMR} and molecular simulations},}\
  }\href@noop {} {\bibfield  {journal} {\bibinfo  {journal} {Science Adv.}\
  }\textbf {\bibinfo {volume} {4}},\ \bibinfo {pages} {eaar8521} (\bibinfo
  {year} {2018})}\BibitemShut {NoStop}%
\bibitem [{\citenamefont {Ml\`ynsk\`y}\ \emph {et~al.}(2020)\citenamefont
  {Ml\`ynsk\`y}, \citenamefont {K\"uhrov\'a}, \citenamefont {K\"uhr},
  \citenamefont {Otyepka}, \citenamefont {Bussi}, \citenamefont {Ban{\'a}{\v
  s}},\ and\ \citenamefont {{\v S}poner}}]{mlynsky2020fine}%
  \BibitemOpen
  \bibfield  {author} {\bibinfo {author} {\bibfnamefont {V.}~\bibnamefont
  {Ml\`ynsk\`y}}, \bibinfo {author} {\bibfnamefont {P.}~\bibnamefont
  {K\"uhrov\'a}}, \bibinfo {author} {\bibfnamefont {T.}~\bibnamefont {K\"uhr}},
  \bibinfo {author} {\bibfnamefont {M.}~\bibnamefont {Otyepka}}, \bibinfo
  {author} {\bibfnamefont {G.}~\bibnamefont {Bussi}}, \bibinfo {author}
  {\bibfnamefont {P.}~\bibnamefont {Ban{\'a}{\v s}}}, \ and\ \bibinfo {author}
  {\bibfnamefont {J.}~\bibnamefont {{\v S}poner}},\ }\bibfield  {title}
  {\enquote {\bibinfo {title} {Fine-tuning of the {AMBER} {RNA} force field
  with a new term adjusting interactions of terminal nucleotides},}\
  }\href@noop {} {\bibfield  {journal} {\bibinfo  {journal} {J. Chem. Theory
  Comput., doi:10.1021/acs.jctc.0c00228}\ } (\bibinfo {year}
  {2020})}\BibitemShut {NoStop}%
\bibitem [{\citenamefont {Mobley}\ \emph {et~al.}(2018)\citenamefont {Mobley},
  \citenamefont {Bannan}, \citenamefont {Rizzi}, \citenamefont {Bayly},
  \citenamefont {Chodera}, \citenamefont {Lim}, \citenamefont {Lim},
  \citenamefont {Beauchamp}, \citenamefont {Slochower}, \citenamefont {Shirts}
  \emph {et~al.}}]{mobley2018escaping}%
  \BibitemOpen
  \bibfield  {author} {\bibinfo {author} {\bibfnamefont {D.~L.}\ \bibnamefont
  {Mobley}}, \bibinfo {author} {\bibfnamefont {C.~C.}\ \bibnamefont {Bannan}},
  \bibinfo {author} {\bibfnamefont {A.}~\bibnamefont {Rizzi}}, \bibinfo
  {author} {\bibfnamefont {C.~I.}\ \bibnamefont {Bayly}}, \bibinfo {author}
  {\bibfnamefont {J.~D.}\ \bibnamefont {Chodera}}, \bibinfo {author}
  {\bibfnamefont {V.~T.}\ \bibnamefont {Lim}}, \bibinfo {author} {\bibfnamefont
  {N.~M.}\ \bibnamefont {Lim}}, \bibinfo {author} {\bibfnamefont {K.~A.}\
  \bibnamefont {Beauchamp}}, \bibinfo {author} {\bibfnamefont {D.~R.}\
  \bibnamefont {Slochower}}, \bibinfo {author} {\bibfnamefont {M.~R.}\
  \bibnamefont {Shirts}},  \emph {et~al.},\ }\bibfield  {title} {\enquote
  {\bibinfo {title} {Escaping atom types in force fields using direct chemical
  perception},}\ }\href@noop {} {\bibfield  {journal} {\bibinfo  {journal} {J.
  Chem. Theory Comput.}\ }\textbf {\bibinfo {volume} {14}},\ \bibinfo {pages}
  {6076--6092} (\bibinfo {year} {2018})}\BibitemShut {NoStop}%
\bibitem [{\citenamefont {Cesari}, \citenamefont {Gil-Ley},\ and\ \citenamefont
  {Bussi}(2016)}]{cesari2016combining}%
  \BibitemOpen
  \bibfield  {author} {\bibinfo {author} {\bibfnamefont {A.}~\bibnamefont
  {Cesari}}, \bibinfo {author} {\bibfnamefont {A.}~\bibnamefont {Gil-Ley}}, \
  and\ \bibinfo {author} {\bibfnamefont {G.}~\bibnamefont {Bussi}},\ }\bibfield
   {title} {\enquote {\bibinfo {title} {Combining simulations and solution
  experiments as a paradigm for {RNA} force field refinement},}\ }\href@noop {}
  {\bibfield  {journal} {\bibinfo  {journal} {J. Chem. Theory Comput.}\
  }\textbf {\bibinfo {volume} {12}},\ \bibinfo {pages} {6192--6200} (\bibinfo
  {year} {2016})}\BibitemShut {NoStop}%
\bibitem [{\citenamefont {Orioli}\ \emph {et~al.}(2020)\citenamefont {Orioli},
  \citenamefont {Larsen}, \citenamefont {Bottaro},\ and\ \citenamefont
  {Lindorff-Larsen}}]{orioli2019learn}%
  \BibitemOpen
  \bibfield  {author} {\bibinfo {author} {\bibfnamefont {S.}~\bibnamefont
  {Orioli}}, \bibinfo {author} {\bibfnamefont {A.~H.}\ \bibnamefont {Larsen}},
  \bibinfo {author} {\bibfnamefont {S.}~\bibnamefont {Bottaro}}, \ and\
  \bibinfo {author} {\bibfnamefont {K.}~\bibnamefont {Lindorff-Larsen}},\
  }\bibfield  {title} {\enquote {\bibinfo {title} {How to learn from
  inconsistencies: Integrating molecular simulations with experimental data},}\
  }\href@noop {} {\bibfield  {journal} {\bibinfo  {journal} {Prog. Mol. Biol.
  Transl.}\ }\textbf {\bibinfo {volume} {170}},\ \bibinfo {pages} {123--176}
  (\bibinfo {year} {2020})}\BibitemShut {NoStop}%
\bibitem [{\citenamefont {Debiec}\ \emph {et~al.}(2016)\citenamefont {Debiec},
  \citenamefont {Cerutti}, \citenamefont {Baker}, \citenamefont {Gronenborn},
  \citenamefont {Case},\ and\ \citenamefont {Chong}}]{debiec2016further}%
  \BibitemOpen
  \bibfield  {author} {\bibinfo {author} {\bibfnamefont {K.~T.}\ \bibnamefont
  {Debiec}}, \bibinfo {author} {\bibfnamefont {D.~S.}\ \bibnamefont {Cerutti}},
  \bibinfo {author} {\bibfnamefont {L.~R.}\ \bibnamefont {Baker}}, \bibinfo
  {author} {\bibfnamefont {A.~M.}\ \bibnamefont {Gronenborn}}, \bibinfo
  {author} {\bibfnamefont {D.~A.}\ \bibnamefont {Case}}, \ and\ \bibinfo
  {author} {\bibfnamefont {L.~T.}\ \bibnamefont {Chong}},\ }\bibfield  {title}
  {\enquote {\bibinfo {title} {Further along the road less traveled: {AMBER}
  ff15ipq, an original protein force field built on a self-consistent physical
  model},}\ }\href@noop {} {\bibfield  {journal} {\bibinfo  {journal} {J. Chem.
  Theory Comput.}\ }\textbf {\bibinfo {volume} {12}},\ \bibinfo {pages}
  {3926--3947} (\bibinfo {year} {2016})}\BibitemShut {NoStop}%
\bibitem [{\citenamefont {K\"uhrov\'a}\ \emph {et~al.}(2019)\citenamefont
  {K\"uhrov\'a}, \citenamefont {Ml\`ynsk\`y}, \citenamefont {Zgarbov{\'a}},
  \citenamefont {Krepl}, \citenamefont {Bussi}, \citenamefont {Best},
  \citenamefont {Otyepka}, \citenamefont {{\v S}poner},\ and\ \citenamefont
  {Ban{\'a}{\v s}}}]{kuhrova2019improving}%
  \BibitemOpen
  \bibfield  {author} {\bibinfo {author} {\bibfnamefont {P.}~\bibnamefont
  {K\"uhrov\'a}}, \bibinfo {author} {\bibfnamefont {V.}~\bibnamefont
  {Ml\`ynsk\`y}}, \bibinfo {author} {\bibfnamefont {M.}~\bibnamefont
  {Zgarbov{\'a}}}, \bibinfo {author} {\bibfnamefont {M.}~\bibnamefont {Krepl}},
  \bibinfo {author} {\bibfnamefont {G.}~\bibnamefont {Bussi}}, \bibinfo
  {author} {\bibfnamefont {R.~B.}\ \bibnamefont {Best}}, \bibinfo {author}
  {\bibfnamefont {M.}~\bibnamefont {Otyepka}}, \bibinfo {author} {\bibfnamefont
  {J.}~\bibnamefont {{\v S}poner}}, \ and\ \bibinfo {author} {\bibfnamefont
  {P.}~\bibnamefont {Ban{\'a}{\v s}}},\ }\bibfield  {title} {\enquote {\bibinfo
  {title} {Improving the performance of the {AMBER} {RNA} force field by tuning
  the hydrogen-bonding interactions},}\ }\href@noop {} {\bibfield  {journal}
  {\bibinfo  {journal} {J. Chem. Theory Comput.}\ }\textbf {\bibinfo {volume}
  {15}},\ \bibinfo {pages} {3288--3305} (\bibinfo {year} {2019})}\BibitemShut
  {NoStop}%
\bibitem [{\citenamefont {Mehta}\ \emph {et~al.}(2019)\citenamefont {Mehta},
  \citenamefont {Bukov}, \citenamefont {Wang}, \citenamefont {Day},
  \citenamefont {Richardson}, \citenamefont {Fisher},\ and\ \citenamefont
  {Schwab}}]{mehta2019ml}%
  \BibitemOpen
  \bibfield  {author} {\bibinfo {author} {\bibfnamefont {P.}~\bibnamefont
  {Mehta}}, \bibinfo {author} {\bibfnamefont {M.}~\bibnamefont {Bukov}},
  \bibinfo {author} {\bibfnamefont {C.-H.}\ \bibnamefont {Wang}}, \bibinfo
  {author} {\bibfnamefont {A.~G.}\ \bibnamefont {Day}}, \bibinfo {author}
  {\bibfnamefont {C.}~\bibnamefont {Richardson}}, \bibinfo {author}
  {\bibfnamefont {C.~K.}\ \bibnamefont {Fisher}}, \ and\ \bibinfo {author}
  {\bibfnamefont {D.~J.}\ \bibnamefont {Schwab}},\ }\bibfield  {title}
  {\enquote {\bibinfo {title} {A high-bias, low-variance introduction to
  machine learning for physicists},}\ }\href {\doibase
  10.1016/j.physrep.2019.03.001} {\bibfield  {journal} {\bibinfo  {journal}
  {Phys. Rep.}\ }\textbf {\bibinfo {volume} {810}},\ \bibinfo {pages} {1–124}
  (\bibinfo {year} {2019})}\BibitemShut {NoStop}%
\bibitem [{\citenamefont {Cawley}\ and\ \citenamefont
  {Talbot}(2010)}]{cawley2010over}%
  \BibitemOpen
  \bibfield  {author} {\bibinfo {author} {\bibfnamefont {G.~C.}\ \bibnamefont
  {Cawley}}\ and\ \bibinfo {author} {\bibfnamefont {N.~L.}\ \bibnamefont
  {Talbot}},\ }\bibfield  {title} {\enquote {\bibinfo {title} {On over-fitting
  in model selection and subsequent selection bias in performance
  evaluation},}\ }\href@noop {} {\bibfield  {journal} {\bibinfo  {journal} {J.
  Mach. Learn. Res.}\ }\textbf {\bibinfo {volume} {11}},\ \bibinfo {pages}
  {2079--2107} (\bibinfo {year} {2010})}\BibitemShut {NoStop}%
\bibitem [{\citenamefont {Rangan}\ \emph {et~al.}(2018)\citenamefont {Rangan},
  \citenamefont {Bonomi}, \citenamefont {Heller}, \citenamefont {Cesari},
  \citenamefont {Bussi},\ and\ \citenamefont
  {Vendruscolo}}]{rangan2018determination}%
  \BibitemOpen
  \bibfield  {author} {\bibinfo {author} {\bibfnamefont {R.}~\bibnamefont
  {Rangan}}, \bibinfo {author} {\bibfnamefont {M.}~\bibnamefont {Bonomi}},
  \bibinfo {author} {\bibfnamefont {G.~T.}\ \bibnamefont {Heller}}, \bibinfo
  {author} {\bibfnamefont {A.}~\bibnamefont {Cesari}}, \bibinfo {author}
  {\bibfnamefont {G.}~\bibnamefont {Bussi}}, \ and\ \bibinfo {author}
  {\bibfnamefont {M.}~\bibnamefont {Vendruscolo}},\ }\bibfield  {title}
  {\enquote {\bibinfo {title} {Determination of structural ensembles of
  proteins: restraining vs reweighting},}\ }\href@noop {} {\bibfield  {journal}
  {\bibinfo  {journal} {J. Chem. Theory Comput.}\ }\textbf {\bibinfo {volume}
  {14}},\ \bibinfo {pages} {6632--6641} (\bibinfo {year} {2018})}\BibitemShut
  {NoStop}%
\bibitem [{\citenamefont {Goodfellow}, \citenamefont {Bengio},\ and\
  \citenamefont {Courville}(2016)}]{goodfellow2016deep}%
  \BibitemOpen
  \bibfield  {author} {\bibinfo {author} {\bibfnamefont {I.}~\bibnamefont
  {Goodfellow}}, \bibinfo {author} {\bibfnamefont {Y.}~\bibnamefont {Bengio}},
  \ and\ \bibinfo {author} {\bibfnamefont {A.}~\bibnamefont {Courville}},\
  }\href@noop {} {\emph {\bibinfo {title} {Deep learning}}}\ (\bibinfo
  {publisher} {MIT press},\ \bibinfo {year} {2016})\BibitemShut {NoStop}%
\bibitem [{\citenamefont {Li}\ and\ \citenamefont
  {Merz~Jr}(2014)}]{li2014taking}%
  \BibitemOpen
  \bibfield  {author} {\bibinfo {author} {\bibfnamefont {P.}~\bibnamefont
  {Li}}\ and\ \bibinfo {author} {\bibfnamefont {K.~M.}\ \bibnamefont
  {Merz~Jr}},\ }\bibfield  {title} {\enquote {\bibinfo {title} {Taking into
  account the ion-induced dipole interaction in the nonbonded model of ions},}\
  }\href@noop {} {\bibfield  {journal} {\bibinfo  {journal} {J. Chem Theory
  Comput.}\ }\textbf {\bibinfo {volume} {10}},\ \bibinfo {pages} {289--297}
  (\bibinfo {year} {2014})}\BibitemShut {NoStop}%
\bibitem [{\citenamefont {Behler}\ and\ \citenamefont
  {Parrinello}(2007)}]{behler2007generalized}%
  \BibitemOpen
  \bibfield  {author} {\bibinfo {author} {\bibfnamefont {J.}~\bibnamefont
  {Behler}}\ and\ \bibinfo {author} {\bibfnamefont {M.}~\bibnamefont
  {Parrinello}},\ }\bibfield  {title} {\enquote {\bibinfo {title} {Generalized
  neural-network representation of high-dimensional potential-energy
  surfaces},}\ }\href@noop {} {\bibfield  {journal} {\bibinfo  {journal} {Phys.
  Rev. Lett.}\ }\textbf {\bibinfo {volume} {98}},\ \bibinfo {pages} {146401}
  (\bibinfo {year} {2007})}\BibitemShut {NoStop}%
\bibitem [{\citenamefont {Smith}, \citenamefont {Isayev},\ and\ \citenamefont
  {Roitberg}(2017)}]{smith2017ani}%
  \BibitemOpen
  \bibfield  {author} {\bibinfo {author} {\bibfnamefont {J.~S.}\ \bibnamefont
  {Smith}}, \bibinfo {author} {\bibfnamefont {O.}~\bibnamefont {Isayev}}, \
  and\ \bibinfo {author} {\bibfnamefont {A.~E.}\ \bibnamefont {Roitberg}},\
  }\bibfield  {title} {\enquote {\bibinfo {title} {{ANI}-1: an extensible
  neural network potential with {DFT} accuracy at force field computational
  cost},}\ }\href@noop {} {\bibfield  {journal} {\bibinfo  {journal} {Chem.
  Sci.}\ }\textbf {\bibinfo {volume} {8}},\ \bibinfo {pages} {3192--3203}
  (\bibinfo {year} {2017})}\BibitemShut {NoStop}%
\bibitem [{\citenamefont {No{\'e}}\ \emph {et~al.}(2020)\citenamefont
  {No{\'e}}, \citenamefont {Tkatchenko}, \citenamefont {M{\"u}ller},\ and\
  \citenamefont {Clementi}}]{noe2020machine}%
  \BibitemOpen
  \bibfield  {author} {\bibinfo {author} {\bibfnamefont {F.}~\bibnamefont
  {No{\'e}}}, \bibinfo {author} {\bibfnamefont {A.}~\bibnamefont {Tkatchenko}},
  \bibinfo {author} {\bibfnamefont {K.-R.}\ \bibnamefont {M{\"u}ller}}, \ and\
  \bibinfo {author} {\bibfnamefont {C.}~\bibnamefont {Clementi}},\ }\bibfield
  {title} {\enquote {\bibinfo {title} {Machine learning for molecular
  simulation},}\ }\href@noop {} {\bibfield  {journal} {\bibinfo  {journal}
  {Annu. Rev. Phys. Chem.}\ }\textbf {\bibinfo {volume} {71}},\ \bibinfo
  {pages} {doi: 10.1146/annurev--physchem--042018--052331} (\bibinfo {year}
  {2020})}\BibitemShut {NoStop}%
\bibitem [{\citenamefont {Gkeka}\ \emph {et~al.}(2020)\citenamefont {Gkeka},
  \citenamefont {Stoltz}, \citenamefont {Farimani}, \citenamefont {Belkacemi},
  \citenamefont {Ceriotti}, \citenamefont {Chodera}, \citenamefont {Dinner},
  \citenamefont {Ferguson}, \citenamefont {Maillet}, \citenamefont {Minoux}
  \emph {et~al.}}]{gkeka2020machine}%
  \BibitemOpen
  \bibfield  {author} {\bibinfo {author} {\bibfnamefont {P.}~\bibnamefont
  {Gkeka}}, \bibinfo {author} {\bibfnamefont {G.}~\bibnamefont {Stoltz}},
  \bibinfo {author} {\bibfnamefont {A.~B.}\ \bibnamefont {Farimani}}, \bibinfo
  {author} {\bibfnamefont {Z.}~\bibnamefont {Belkacemi}}, \bibinfo {author}
  {\bibfnamefont {M.}~\bibnamefont {Ceriotti}}, \bibinfo {author}
  {\bibfnamefont {J.}~\bibnamefont {Chodera}}, \bibinfo {author} {\bibfnamefont
  {A.~R.}\ \bibnamefont {Dinner}}, \bibinfo {author} {\bibfnamefont
  {A.}~\bibnamefont {Ferguson}}, \bibinfo {author} {\bibfnamefont {J.-B.}\
  \bibnamefont {Maillet}}, \bibinfo {author} {\bibfnamefont {H.}~\bibnamefont
  {Minoux}},  \emph {et~al.},\ }\bibfield  {title} {\enquote {\bibinfo {title}
  {Machine learning force fields and coarse-grained variables in molecular
  dynamics: application to materials and biological systems},}\ }\href@noop {}
  {\bibfield  {journal} {\bibinfo  {journal} {arXiv preprint arXiv:2004.06950}\
  } (\bibinfo {year} {2020})}\BibitemShut {NoStop}%
\end{thebibliography}%

\end{document}